\documentclass[twocolumn,final,longtitle]{elsarticle}

\usepackage{framed,multirow}

\usepackage{amssymb}
\usepackage{latexsym}
\usepackage{upgreek}
\usepackage[ruled,linesnumbered]{algorithm2e}
\usepackage{multirow}
\usepackage{romannum}
\usepackage{enumerate}
\usepackage{amsmath}
\usepackage{natbib}
\usepackage{url}
\usepackage{xcolor}
\usepackage[counterclock]{rotating}
\usepackage{hyperref}
\usepackage{adjustbox}
\definecolor{newcolor}{rgb}{.8,.349,.1}

\journal{Medical Image Analysis}

\begin{document}

\begin{frontmatter}

\title{Combiner and HyperCombiner Networks: Rules to Combine Multimodality MR Images for Prostate Cancer Localisation }


\author[1,2]{{Wen} {Yan}\corref{cor1}} \cortext[cor1]{Corresponding author}\ead{wenyan6-c@my.cityu.edu.hk}\fnref{fn1}\fntext[fn1]{submitted to Medical Image Analysis: 28 September 2022; accepted: 13 November 2023; available: 21 November 2023}
\author[5,1]{{Bernard} {Chiu}}\ead{bchiu@wlu.ca}

\author[2]{ {Ziyi} {Shen}}\ead{ziyi-shen@ucl.ac.uk}
\author[2]{{Qianye} {Yang}} \ead{qianye.yang.19@ucl.ac.uk}
\author[3]{{Tom} {Syer}}\ead{t.syer@ucl.ac.uk}
\author[2]{{Zhe} {Min}}\ead{z.min@ucl.ac.uk} 
\author[3]{{Shonit} {Punwani}}\ead{s.punwani@ucl.ac.uk}
\author[4]{{Mark} {Emberton}}\ead{m.emberton@ucl.ac.uk}
\author[3]{{David} {Atkinson}}\ead{d.atkinson@ucl.ac.uk}
\author[2]{{Dean C.} {Barratt}}\ead{d.barratt@ucl.ac.uk}
\author[2]{{Yipeng} {Hu}}\ead{yipeng.hu@ucl.ac.uk}

\address[1]{Department of Electrical Engineering, City University of Hong Kong, 83 Tat Chee Avenue, Hong Kong, China}

\address[2]{Centre for Medical Image Computing; Department of Medical Physics \& Biomedical Engineering; Wellcome/EPSRC Centre for Interventional and Surgical Sciences, University College London, Gower St, London WC1E 6BT, London, U.K.}

\address[3]{Centre for Medical Imaging, Division of Medicine, University College London, London W1W 7TS, U.K.}

\address[4]{Division of Surgery \& Interventional Science, University College London, Gower St, London WC1E 6BT, London, U.K.}

\address[5]{Department of Physics \& Computer Science, Wilfrid Laurier University, 75 University Avenue West Waterloo, Ontario N2L 3C5, Canada}

\begin{abstract}
One of the distinct characteristics of radiologists reading multiparametric prostate MR scans, using reporting systems like PI-RADS v2.1, is to score individual types of MR modalities, including T2-weighted, diffusion-weighted, and dynamic contrast-enhanced, and then combine these image-modality-specific scores using standardised decision rules to predict the likelihood of clinically significant cancer. This work aims to demonstrate that it is feasible for low-dimensional parametric models to model such decision rules in the proposed Combiner networks, without compromising the accuracy of predicting radiologic labels. 
First, we demonstrate that either a linear mixture model or a nonlinear stacking model is sufficient to model PI-RADS decision rules for localising prostate cancer. Second, parameters of these combining models are proposed as hyperparameters, weighing independent representations of individual image modalities in the Combiner network training, as opposed to end-to-end modality ensemble. A HyperCombiner network is developed to train a single image segmentation network that can be conditioned on these hyperparameters during inference for much-improved efficiency. Experimental results based on 751 cases from 651 patients compare the proposed rule-modelling approaches with other commonly-adopted end-to-end networks, in this downstream application of automating radiologist labelling on multiparametric MR. By acquiring and interpreting the modality combining rules, specifically the linear-weights or odds ratios associated with individual image modalities, three clinical applications are quantitatively presented and contextualised in the prostate cancer segmentation application, including modality availability assessment, importance quantification and rule discovery. 
\end{abstract}


\end{frontmatter}

 \section{Introduction}
\label{sec:introduction}
\begin{figure}
\caption{The illustration of PI-RADS v2 scoring system using a scale of 1--5 (left) and an example of the derived binary classification system used in this study (right).\label{fig:pirads}}
\includegraphics[width=\linewidth]{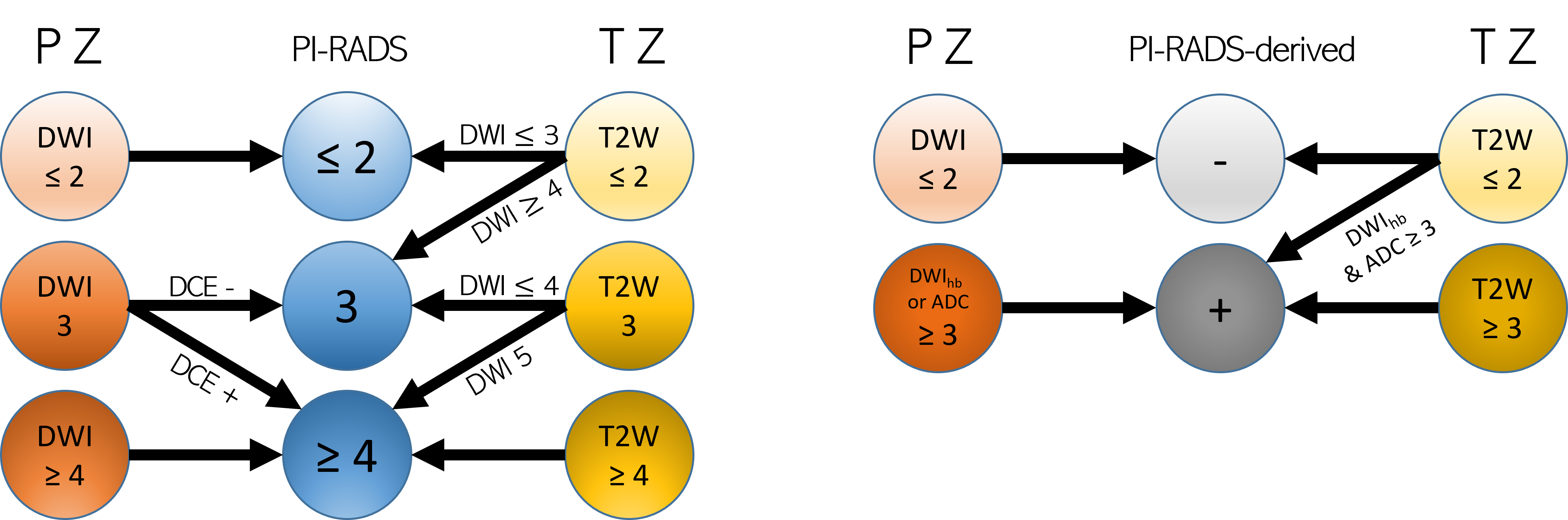}
\end{figure}

The use of multiparametric magnetic resonance (mpMR) images, including
T2-weighted (T2W), diffusion-weighted imaging (DWI) with its high b-value
weighting (DWI$_{hb}$) and apparent diffusion coefficient (ADC) maps, along
with dynamic contrast-enhanced (DCE) modalities, is recommended by various
national guidelines to reduce unnecessary invasive biopsies and treatment in
patients with suspected prostate
cancer~\citep{ahmed2017diagnostic,turkbey2011multiparametric,haider2007combined}.
One of these guidelines is the Prostate Imaging Reporting and Data System
(PI-RADS), both V1~\citep{rosenkrantz2013prostate} and
V2~\citep{weinreb2016pi}, which was developed to standardise the mpMR
assessment process and improve agreement among radiologists using a detailed
scoring criterion.
According to the five-point PI-RADS system, DWI is the primary modality for
evaluating lesions in the peripheral zone (PZ), while DCE can upgrade lesions
with scores of 3 to 4. In transition zone (TZ), the overall assessment
primarily relies on the T2W score, while a lesion with a score of 2 or 3 on
T2W may be raised to 3 or 4, respectively, based on additional evidence on
DWI~\citep{weinreb2016pi}, as illustrated in \ref{fig:pirads} (left).
Reporting mpMR images is an expertise-demanding and labour-intensive task for
radiologists. Much literature
~\citep{litjens2014computer,schelb2019classification,chen2020automatic,hambarde2020prostate,chiou2020harnessing,qian2021procdet}
has demonstrated that machine learning approaches, including deep learning, are
potentially valuable tools that can help radiologists as a first- or second
assisting reader. Existing machine learning methods have exploited radiology
knowledge to improve either classification or segmentation accuracy, including
utilising prior knowledge of different lesion types in prostate zonal
anatomy~\citep{giannini2015fully,duran2020prostate,van2021deep} and image
modalities~\citep{de2020deep,chiou2020harnessing,bonekamp2018radiomic} from available MR modalities\footnote{In this work, an image modality is refereed to a 3D image volume that observers commonly visualise, such as T2W, ADC, DWI$_{hb}$ or DCE, not necessarily corresponding to different imaging-physics-defined MR sequences.}

This paper investigates a different set of questions, {other than improving the segmentation performance}: (a) whether summary information obtained from individual image modalities, such as ``positive on T2W'' or ``anterior lesion on ADC'' commonly reported in clinical practice, can first be predicted by deep learning models, before subsequent analysis? and (b) How and whether this image modality-independent representation can achieve adequate model performance, compared to the above end-to-end representation learning approaches, but for better model interpretation and feedback, which may lead to {an improvement to the existing combining rules such as PI-RADS}? 
These questions are intriguing because the ability to quantify the importance, contribution and necessity of individual image modalities to the final prediction provides insights into interpretation at the image modality level. This paper discusses a set of specific clinical applications enabled by this individual modality quantification, in \ref{sec:method}, which is also related to the literature discussing interpretable machine learning, further discussed in \ref{sec:experiment}.

\subsection*{Combining as late as possible, with hyperparameters}

Machine learning models can use a single image modality as input to ensure their predictions are independent of other image modalities. 
 {This independence reflects what is recommended by the PI-RADS, based on available clinical evidence, for streamlining the interpretation of mpMR images. It is designed for first reading a predominant image type, and subsequently taking other types of images for improved staging.}
Similarly, the decision process of combining these modality-independent predictions, e.g.,~lesion-level classification and voxel-level segmentation, can be modelled by machine learning models in non-parametric forms such as averaging or majority voting. 

Notably, model-combining methods, such as ensembles and committees, have been extensively studied in the literature to improve the overall performance of the combined model. The model combining methods, in turn, can be parameterised with learnable parameters as latent variables in an end-to-end optimisation process. Examples of such parameters include the mixing coefficients in mixture models and the posterior probabilities in Bayesian model averaging. Some relevant examples are included in \ref{sec:intro.related}.

\begin{figure*}
\caption{Graphical models from (a) to (c) use a single (shared) network with parameters $\uptheta$, where $\mathbf{y}^1$, $\mathbf{y}^2$ and $\mathbf{y}$ are intermediate features and $\upalpha$ is a set of parameters, to combine the inputs $\mathbf{x}^1$ and $\mathbf{x}^2$ and output $\mathbf{z}$. (a) an example of late combiner models, of interest in this study, with observed $\upalpha$ in shaded circle, (b) a late fusion in combining methods with learnable $\upalpha$, (c) an early fusion model for comparison. Hollow circles and solid dots indicate random variables and deterministic parameters, respectively. This is one example to show the difference between the combiner models (a) and other model combining methods (b and c), among other possible probabilistic graphical representations. Besides, (d) and (e) are schematic diagrams of the proposed Combiner and HyperCombiner, respectively, with three image modalities as example inputs.\label{fig:graphical}}
\includegraphics[width=\linewidth]{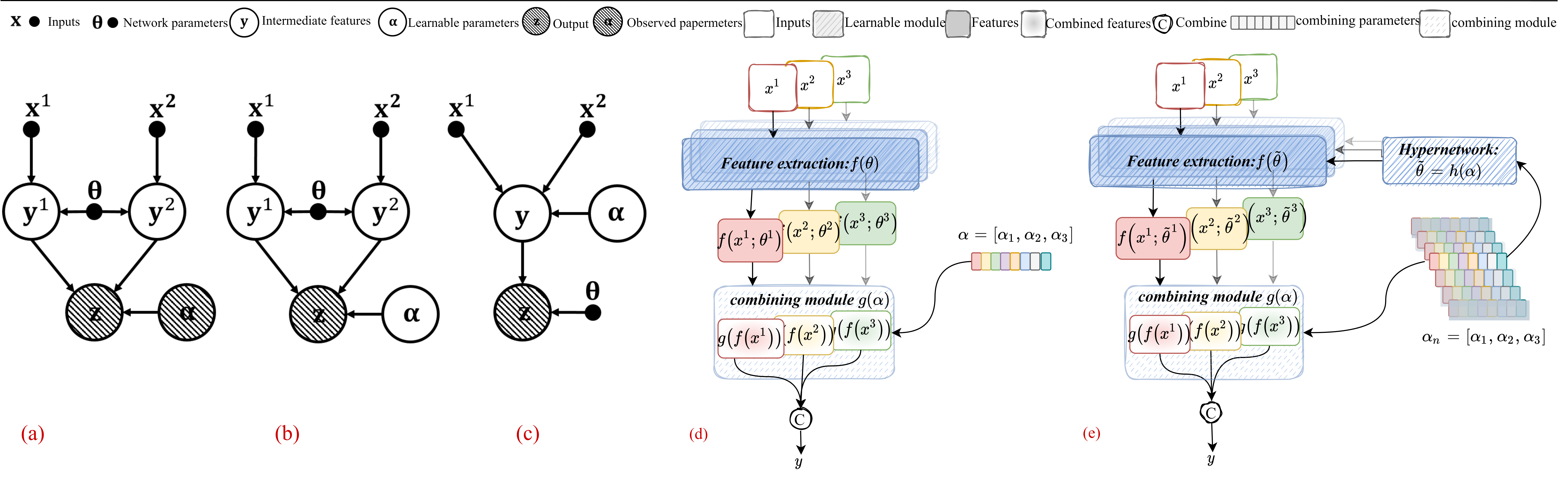}
\end{figure*}

In contrast, this work combines predictions using decision-making processes that are ``hyperparameter-ised'' in order to enable a set of interesting applications, including (1) assessing existing decision processes in combining image modalities, and (2) discovering new decision rules with varying practical constraints, such as missing modality or known low-quality or otherwise challenging image modalities.
As illustrated by the graphical model in \ref{fig:graphical}(a)--(b), model (a) represents the hyperprameterised combining model, referred to as ``combiner models'' in this paper, which incorporates a set of predefined hyperparameters to combine the individual image modalities. On the other hand, model (b) represents a combining method that jointly optimises the representation of each of the individual image modalities and the combining decision with learnable parameters, denoted as the ``late fusion''. The alternative ``early fusion'', as illustrated in \ref{fig:graphical}(c), could also be used to assess the importance and availability of individual modalities. However, combining inputs before representation learning loses the direct influence on the final prediction due to the less interpretable ``black-boxed'' neural networks. The direct connection from individual image representation to the final prediction, as depicted in \ref{fig:graphical}(a) and (b), plays a significant role in motivating and enabling the aforementioned applications related to modality comparison and rule discovery. This connection is often established through a weighted sum or a more general form of aggregation.

This paper describes the development of ``Combiner'' networks based on individual modalities of mpMR images in conjunction with hyperparameterised decision-making models, as shown in \ref{fig:graphical}(d). Two rule-combining methods are proposed to represent linear and nonlinear decision rules using low dimensional hyperparameters. 
In addition, a hypernetwork is also developed, as shown in \ref{fig:graphical}(e), which incorporates varying hyperparameter values within a single neural network. The purpose of this hypernetwork is to enhance the practicality and efficiency of the proposed combiner networks shown in \ref{fig:graphical}(d), as well as facilitate the search for optimum rules.
When there is a need to adjust hyperparameters for analysis purposes, as is the case in this study, hypernetwork is used to enable hyperparameter adjustment during inference. This approach eliminates the need to train many models with different hyperparameter values separately. These hypernetworks are referred to as ``HyperCombiner'' networks in this work.
 {We would like to emphasise that the primary objective is to propose a new approach for modelling combining rules based on modality-independent predictions. The focus of this study is not on improving segmentation performance but on developing the rule-modelling approach and exploring its interpretability within the context of the segmentation application.}
The main hypothesis investigated in this study is that, with the added rule modelling, lesions segmentation models can achieve a non-compromising performance but are more interpretable. Experimental results are presented based on clinical mpMR images from 651 prostate cancer patients and multiple radiologist reports obtained in five clinical studies. 

This study then investigates the benefit of incorporating the rule modelling, in Sec.\ref{sec:experiment} and \ref{sec:results}, which enables addressing the following specific clinical questions. 

\begin{itemize}
 \item Does separating the representation for individual image modalities sacrifice the model performance, compared with end-to-end learning?
 \item Are hypernetworks capable of summarising different hyperparameter values, representing different combining rules, in a single neural network?
 \item What is the impact of individual image modalities on the model performance, by varying hyperparameters?
 \item Can the hyperparameters be optimised to obtain optimum decision rules via hyperparameter search?
 \item Are the empirical PI-RADS-derived rules better or worse than the above numerically optimised decision rules? 
 \item Would conclusion differ for tumours in different zones? 
\end{itemize}

\section{Related Work}

\label{sec:intro.related}

\subsection{Machine Learning for Prostate {mpMR} images}

Recent studies have shown that mpMR images provide better indications of clinically significant
prostate cancer compared to a single type of MR images
~\citep{chan2003detection,haider2007combined}. Various methods have been
proposed to detect and localise prostate lesions on the mpMR images. 
Weak supervision using patient-level labels has also been investigated to avoid
the need for lesion annotations~\citep{mehta2021computer,saha2021end}. Others
learn such features using labels from radiologists or histopathology results,
albeit the latter of which often are based on targeted
biopsy~\citep{litjens2014computer}\citep{schelb2019classification,chen2020automatic,mehralivand2022cascaded,eidex2022mri}.~\citet{litjens2014computer}
proposed a cascaded lesion detection system and achieved a patient-level ROC
AUC of 0.83 for detecting clinically significant cancer on a data cohort
including 347 patients.~\citet{mehralivand2022cascaded} presented an approach
for automated detection and segmentation of prostate MRI lesions, followed by a
4-class classification task to predict PI-RADS categories 2 to 5. The
experiments were conducted on 1043 MR scans and yielded a 0.359 median DSC for
lesion segmentation and a 0.308 overall PI-RADS classification accuracy.
 ~\citet{eidex2022mri} presented a cascaded network to
segment the prostate and dominant intraprostatic lesions (DIL). A high
segmentation accuracy is reported, with Dice scores of 0.896 and 0.843 for
prostate and DIL segmentation, respectively.
Besides, end-to-end networks have been proposed to segment suspected lesions on
mpMR images
~\citep{chen2020automatic,de2020deep,duran2020prostate,duran2022prostattention,antonelli2019machine,hambarde2020prostate,chiou2020harnessing,qian2021procdet,woznicki2020multiparametric}. 

Prior knowledge of radiology, including clinical features and zonal prostate
anatomy, is often used to improve the accuracy of lesion
segmentation.~\citet{chen2020automatic} used multi-branched UNet to extract
features from different modalities of MR image, achieving a Dice similarity
coefficient of 0.72 and a sensitivity of 0.74 at 1.00 specificity.
~\citet{de2020deep} presented a method to segment prostate lesions by encoding
the ordinal Gleason groups, which scored a voxel-wise weighted kappa of 0.446~$\pm$~0.082 and a DSC of 0.370~$\pm$~0.046.~\citet{duran2022prostattention}
proposed an attention CNN for joint multi-class segmentation of prostate and
cancer lesions by Gleason scores, which outperformed well-tuned UNet, Attention
UNet, ENet and DeepLabv3$+$ at detecting clinically significant
prostate cancers.~\citet{chiou2020harnessing} proposed a domain adaption
approach from routine mpMR images to VERDICT, a customised MR modality for
prostate cancer, which significantly improved segmentation.

In addition, literature has investigated the relationship between the
assessment using PI-RADS by radiologists and the automated assessment by algorithms~\citep{schelb2019classification,youn2021detection,wang2017machine,sanford2020deep}.~\citet{wang2017machine}
verified in 54 study cohorts that machine learning could improve the
performance of PI-RADS assessment.~\citet{youn2021detection} conducted
experiments on datasets including patients who underwent pre-biopsy MRI and
prostate biopsy to compare lesion detection and PI-RADS classification
performance between UNet, clinical reports and radiologists. The results showed
that algorithms achieved moderate diagnostic performance with experts in
PI-RADS and were similar to clinical reports from various radiologists in
clinical practice.

Over the last few years, much work has been published to address the problem of prostate cancer localisation on mpMR images using machine learning-based methods, focusing on the performance of segmentation or detection. However, little has addressed the interpretable image modality-related questions posed at the end of \ref{sec:introduction}.

\subsection{Model Combining}

In literature, the term \textit{model combining} refers to mixing or fusing individual branches of a model as an ensemble, often aiming to improve the performance of a combined model and is widely used in medical imaging tasks. Although they have a different aim from the \textit{combiner} models of interest in this work, discussed in \ref{sec:introduction}, many similarities in methodologies can be shared between the two.

\citet{greenspan2006constrained} presented an unsupervised Gaussian mixture
model (GMM)-based segmentation method for tissue classification of MR images of
the brain.~\citet{hassan2019robust} also proposed an approach that employed
fuzzy intelligence and GMM to detect carotid artery plaque in ultrasound images
with high precision and strong robustness to noise.~\citet{nguyen2012fast}
proposed Markov random field with GMM that directly applied the EM algorithm
for optimisation.

The two terms, early fusion and late fusion, are largely used in model
combining, e.g.~for multi-modality data
input~\citep{d2015review,snoek2005early,bi2022hyper}. As illustrated in
\ref{fig:graphical} and discussed in \ref{sec:introduction},
the late combiner developed in this work is conceptually close to the late
fusion.~\citet{albashish2015multi} proposed a prostate cancer diagnosis method
that fused the probabilities outputs from the support vector machine and
recursive feature elimination classifiers by a summary rule. 
\citet{wang2021modeling} proposed a fusion method for lung cancer survival
analysis that exploited the correlations among predictions produced by
different modalities regarding the model uncertainty and fused them by a
weighted average scheme.~\citet{ghoniem2021multi} developed a multi-modal
fusion framework for ovarian cancer diagnosis, implementing CNN and LSTM as the
base classifiers and weighted the classification outputs.~\citet{trong2020late}
developed a classification approach for weeds classification by using the late
fusion of multi-modal networks via a voting method. 

The works mentioned above combine base learners' scores by applying fusion
rules, such as voting, averaging and summation, while others search for deep
fusion methods based on optimised strategies.~\citet{boulahia2021early}
designed an end-to-end late fusion network where a deep neural network computes
the merging score for action recognition.~\citet{mehta2021computer} utilised a
two-level support vector machine (SVM) to fuse the patient probability of
prostate cancer predicted from mpMR features and clinical features. 
These fusion methods are also referred to as stacking ensemble learning
~\citep{ayache2007classifier}, which uses a meta-learner to combine the
predictions of base
learners~\citep{wolpert1992stacked}.~\citet{ksikazek2020development} proposed a
stacking learning (ensemble) method to detect hepatocellular carcinoma, using
SVM classifier to fuse seven base classifiers, including K-nearest neighbour,
random forest, and Na{\"\i}ve Bayes. 
\citet{taspinar2021classification} used three single base learners, including
the SVM, logistics regression, and artificial neural network models, to produce
classification probabilities and fuse them by a stacking meta-model to classify
a total of 3486 chest X-ray images into three classes.~\citet{wang2019stacking}
utilised a random forest classifier-based stacking technique to simultaneously
construct the diagnostic model and combine them by decision tree for prostate
cancer detection.~\citet{saha2021end} used a decision fusion model to ensemble
the patch-based and image-based prediction by SVM classifiers to reduce false
positives on prostate lesion segmentation.

In summary, these model combining and fusion methods have been proposed to
improve the generalisation of models or methods based on data from multiple
modalities, with promising results. In contrast to the alternative approach of
end-to-end black-box representation learning, they demonstrated that such
human-adopted practice in combining data- or method-diverged decisions might be
advantageous in practical applications, which are often constrained by data and
other resources. {However, we argue that the focus on
performance alone overlooks the potential benefits of leveraging the simplicity
and interpretability of these model-combining methods. Additionally, there is a
lack of methodology and application that employ the concept of the so-called
``interpretability''~\citep{molnar2020interpretable}, which provide valuable
insights and understanding of the decision-making process.}

\subsection{Hypernetworks}

Deep learning methods in practice depend on tuning hyperparameters, such as the
weights of the regularisation terms, that can significantly affect performance.
Therefore, such hyperparameters must be carefully tuned to achieve the best
result, which requires considerable time and computing resources. To address
this issue, hypernetworks are proposed to train neural networks automatically
adapted to various hyperparameters by re-parameterising the main network as a
function of the hyperparameters, using a small auxiliary network.
~\citet{ha2016hypernetworks} ~and~\citet{DBLP:journals/corr/abs-1802-09419} gave a
theoretical justification for hypernetworks and used a gradient-based
optimisation to tune hyperparameters. 
\citet{wang2021regularization} proposed a reconstruction network that is
independent of regularisation methods. In this approach, the parameters of the
reconstruction network were generated by a hypernetwork, which functions based
on the regularisation weights.~\citet{hoopes2021hypermorph} proposed a
learning-based strategy for deformable image registration, which replaces the
tedious process of tuning important registration hyperparameters. In this
approach, the weights of the registration network were generated based on the
hyperparameter values, which were then utilised in the loss function.

Others used hypernetwork training to adapt to multiple tasks using only a
single hypernetwork.~\citet{brock2017smash} accelerated the training procedure
by learning an auxiliary hypernetwork that generates the weights of the main
model for various architectures in a single training.
~\citet{klocek2019hypernetwork} constructed an image representation network
that takes an input image and returns weights to the target network to map
points from the plane into their corresponding colours in the image. 

Despite the promising studies mentioned above, hypernetworks have rarely been used to explain the ``black box'' of neural networks, which remains a topic to be explored.

\section{Method}
\label{sec:method}
\subsection{Image Modality-Independent Base Networks}

\label{sec:method.single-net}
Let us consider a problem to segment regions of interest from $\mathbf\uptau$ types of images $\mathbf{X}^{\tau}$ for each subject.\footnote{Uppercase letters X, Y, Z and $\mathcal{C}$ denote random variables (if in bold, random vectors), with the corresponding lowercase letters are the observed scalars and vectors.} The segmentation using individual image modalities can be formulated as a voxel classification, such that each voxel $i, i=1, \ldots ,I$ is classified as either lesion or non-lesion voxel, where $I$ is the number of voxels in each image. We would like to train a neural network $f^{\tau}$ with a set of parameters $\mathbf\uptheta^{\tau}$ for each image modality $\tau$ to predict the class probabilities $\mathbf{Y}^{\tau}=[Y^{\tau}_{1}, \ldots ,Y^{\tau}_{I}]^{\top}, Y^{\tau}_{i} \in [0,1], \tau=1, \ldots ,\mathbf\uptau$:
\begin{equation}
\label{eq:network_type}
  \mathbf{Y}^{\tau}=f^{\tau}(\mathbf{X}^{\tau}; \mathbf\uptheta^{\tau})
\end{equation}

Given labelled training images $\mathbf{x}^{\tau}_{j}=[x^{\tau}_{1,j}, \ldots ,x^{\tau}_{i,j}]^{\top}$ from $J$ subjects, the networks $f^{\tau}$ can be trained independently by minimising a segmentation loss $\mathcal{L}$ between the predicted class probabilities $\mathbf{y}^{\tau}_{j}=[y^{\tau}_{1,j}, \ldots ,y^{\tau}_{i,j}]^{\top}$ and the subject-specific labels $\mathbf{t}_{j}=[t_{1,j}, \ldots ,t_{i,j}]^{\top}$, i.e.,~the ground-truth segmentation masks available for individual subjects, where $j=1, \ldots ,J$.
\begin{equation}
\label{eq:base-opt}
\hat{\mathbf\uptheta}^{\tau} = arg\min_{\mathbf\uptheta} \sum_{j=1}^{J}\mathcal{L}(\mathbf{y}^{\tau}_{j}, \mathbf{t}_{j})
\end{equation}
where $\mathbf{y}^{\tau}_{j}=f^{\tau}(\mathbf{x}^{\tau}_{j}; \mathbf\uptheta^{\tau})$ and $\hat{\mathbf\uptheta}^{\tau}$ is the set of optimised network parameters.

In our application, we consider a binary classification denoting all identified lesions as positive with $\mathbf\uptau=3$ types of images, $y^{\tau}_{i}=\{0,1\}$ and $\tau={1,2,3}$ for respective T2W, DWI$_{hb}$ and ADC images. The formulation described here may be generalised to multi-class classification and more image types when, for example, detailed radiological/histopathological grading, such as the five-point PI-RADS scores and other image modalities, including DCE, are available. A combination of cross-entropy and a soft Dice loss is used in this study.
\begin{equation}
\label{eq:loss}
\begin{aligned}
\mathcal{L}(\mathbf{y}^{\tau}_{j}, \mathbf{t}_{j}) = \sum_{i=1}^{I}[t_{i,j}\log{y^{\tau}_{i,j}}+(1-t_{i,j})\log{(1-y^{\tau}_{i,j})}] \\
 \ - \ \frac{2 \sum_{i=1}^{I}(y^{\tau}_{i,j} \cdot t^{\tau}_{i,j})}{\sum_{i=1}^{I}y^{\tau}_{i,j} + \sum_{i=1}^{I}t^{\tau}_{i,j}}
\end{aligned}
\end{equation}
These image modality-specific networks are hereinafter referred to as ``Base networks''.

\subsection{Modelling Rules to Combine: {Combiner} Networks}

\label{sec:method.combine-net}
With the image modality-independent predictions $\mathbf{Y}^{\tau}$ with the now shared $\mathbf\uptheta$ represents the feature extractor in the ``Combiner network'':
\begin{equation}
\label{eq:network_shared}
  \mathbf{Y}^{\tau}=f(\mathbf{X}^{\tau}; \mathbf\uptheta).
\end{equation} 

\begin{figure*}
\caption{Illustration of the proposed Combiner (left) and HyperCombiner networks (right), which receive three types of images, T2W, DWI$_{hd}$ and ADC, as inputs. The hyperparameters $\mathbf\upalpha$ are generated conditionally on certain combination rules (b) and are used as combination parameters to combine the outputs of the three images in training. The red parameters in Combiner and HyperCombiner are trainable weights, while all weights $\mathbf\uptheta$ in HyperCombiner are non-trainable and are generated by an auxiliary hypernetwork $\tilde{\mathbf\uptheta}=h(\mathbf\upalpha;\mathbf\upphi)$.\label{fig:net}}
\includegraphics[width=\linewidth]{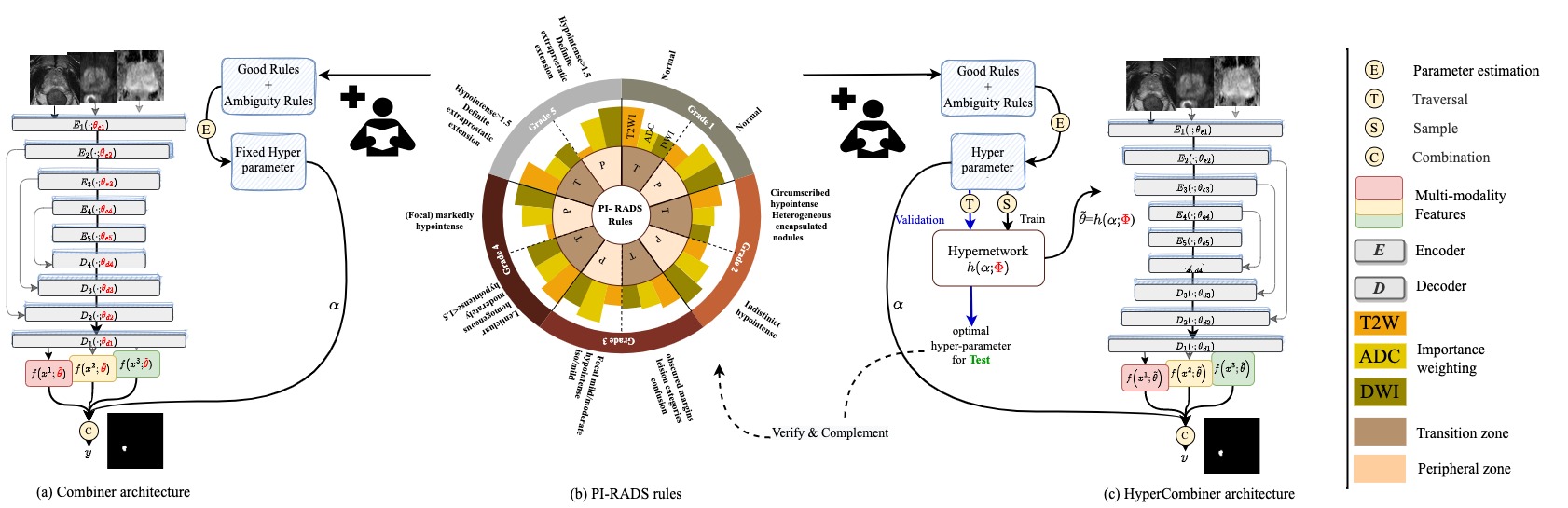}
\end{figure*}

This section describes how these predictions can be combined into a
``consensus'' segmentation for each subject, as shown in \ref{fig:net}
(left). We share weights $\mathbf\uptheta$ among feature extractor $f(\mathbf{X}^{\tau};\mathbf\uptheta)$ in
Combiner with respect to three modalities of images, because a shared network
may benefit from the multi-task effect, where different tasks regularise each
other~\citep{crawshaw2020multi}. However, separately trained networks may also
be sufficient in this application, and further investigation is done in
\ref{sec:results-comb}.

We propose two parametric functions, a linear mixture model and a nonlinear stacking model, before describing how these two functions can themselves be optimised and used to combine the base segmentation networks.

These proposed methods are closely related to several machine learning
techniques, such as stacking ensemble~\citep{wang2019stacking}, gating
~\citep{zhou2020gfnet} and late fusion in multi-task learning
~\citep{wang2021modeling}, which are discussed further in
\ref{sec:discussion}.

\subsubsection{Linear mixture model}
\label{sec:method.combine-net.linear}
The linear model $g^{lin}$ is a weighted sum of predictions over three image modalities, therefore a voxel-wise function of $\mathcal{Y}=[\mathbf{Y}^{1}, \ldots ,\mathbf{Y}^{\tau}]$:
\begin{equation}
\label{eq:linear}
  Z=g^{lin}(\mathcal{Y};\mathbf\upalpha)=\sum_{\tau=1}^{\mathbf\uptau}\alpha_{\tau} \cdot \mathbf{Y}^{\tau}=\sum_{\tau=1}^{\mathbf\uptau}\alpha_{\tau} \cdot f(\mathbf{X}^{\tau};\mathbf\uptheta)
\end{equation}
where $\mathbf\upalpha=[\alpha_1, \ldots ,\alpha_{\mathbf\uptau}]^{\top}$, $\alpha_{\tau} \in [0,1]$ and $\sum_{\tau=1}^{\mathbf\uptau}\alpha_{\tau}=1$ is a set of ``mixing parameters'', as in a mixture model. Recall that image modality-specific class probability $Y^{\tau}_{i}$ represents a conditional probability given the image modality $p(\mathcal{C}\mid\tau)$, where $Y^{\tau}_{i}=p(\mathcal{C}\mid\tau)$. Thus, the mixture model represents a joint probability over all possible image modalities, $Z_{i}=p(\mathcal{C})=\sum_{\tau=1}^{\mathbf\uptau}p(\tau)p(\mathcal{C}\mid\tau)$
where $\mathcal{C}$ denotes the class(es) of interest and $p(\tau)=\alpha_{\tau}$.

\subsubsection{Nonlinear stacking model}
\label{sec:method.combine-net.nonlinear}
Combining the predictions using a nonlinear function may also be useful. For example, the maximum function representing majority voting can be approximated by a sigmoid function or its variants. This section describes a generalisation to the above linear mixture model allowing such a nonlinear combination of class probabilities, such that:
\begin{equation}
\label{eq:nonlinear}
  Z=g^{nonl}(\mathcal{Y};\mathbf\upbeta) = \sigma(\sum_{\tau=1}^{\mathbf\uptau}\beta_{\tau} \cdot \mathbf{Y}^{\tau}+\beta_0)=\sigma[\sum_{\tau=1}^{\mathbf\uptau}\beta_{\tau} \cdot f(\mathbf{X}^{\tau};\mathbf\uptheta)+\beta_0]
\end{equation}
where $\mathbf\upbeta=[\beta_0,\beta_1, \ldots ,\beta_{\mathbf\uptau}]^{\top}$ is a set of ``stacking parameters'' and $\beta_0$ is the bias term. For binary classification in this study, the logistic sigmoid function $\sigma(a)=(1+e^{-a})^{-1}$ is used to represent class probability $Z_{i}=p(\mathcal{C}) \in [0,1]$. This alternative nonlinear stacking function approximates the integral of $p(\mathcal{C}\mid\tau)$, over all images types $\tau$, to obtain the marginalised class probability $p(\mathcal{C})=\int_{\tau}p(\mathcal{C}\mid\tau) \cdot d\tau$. Although the mixture model can be considered as a special case of this more general stacking formulation, these two terms are distinguished in the paper for clarity.

\subsubsection{Training hyperparameterised Combiner networks}

\label{sec:method.combine-net.hyperparameters}
This section describes the segmentation network that can utilise all image types, as opposed to the Base networks. This single network can be trained using either the mixing parameters $\mathbf\upalpha$ or the stacking parameters $\mathbf\upbeta$ as the \textit{network hyperparameters}. 

Now, given all $\mathbf\uptau$ types of images $\mathbf{x}^{\tau}_{j}$ and labels $\mathbf{t}_{j}$ for each of the available training $J$ subjects, as in \ref{sec:method.single-net}, the network can be optimised by minimising the same loss as in Eq.~\eqref{eq:loss}:
\begin{equation}
\label{eq:join-opt}
\hat{\mathbf\uptheta} = arg\min_{\mathbf\uptheta} \sum_{j=1}^{J}\mathcal{L}(\mathbf{z}_{j}, \mathbf{t}_{j})
\end{equation}
where $\mathbf{z}_{j}=[z_{i,j}, \ldots ,z_{i,j}]^{\top}$ and $z_{i,j}$ can be obtained through either $g^{lin}([y_{i,j}^{1}, \ldots ,y_{i,j}^{\tau}]^{\top};\mathbf\upalpha)$ or $g^{nonl}([y_{i,j}^{1}, \ldots ,y_{i,j}^{\tau}]^{\top};\mathbf\upbeta)$ in Eq.~\eqref{eq:linear} or Eq.~\eqref{eq:nonlinear}, respectively. $\mathbf\uptheta$ is a set of network parameters that are no longer specific to any image modality, neither is the optimal $\hat{\mathbf\uptheta}$.

It is worth noting here that if $\mathbf\upalpha$ or $\mathbf\upbeta$ were to be trained simultaneously with the shared network parameters $\mathbf\uptheta$, they are no longer hyperparameters. 
As discussed in \ref{sec:introduction}, previous work has suggested that a conceptually similar multi-branch formulation may provide potential performance benefits through architectural constraints or additional image modality-specific supervision. However, these methods that are promising for improving generalisation are considered beyond the focus of this work and are not discussed further. 

\subsection{ {HyperCombiner} Networks}
\label{sec:method.hyper-net}
This section introduces an additional Combiner with various hyperparameters, as shown in \ref{fig:net} (right). A single trained network can be conditioned on varying hyperparameters during inference, instead of training individual networks for each different set of hyperparameter values.
Let a hypernetwork module be a parametric function of a set of hyperparameters, e.g.,~$\mathbf\upalpha$:
\begin{equation}
  \tilde{\mathbf\uptheta}=h(\mathbf\upalpha;\mathbf\upphi)
\end{equation}
where $\mathbf\upphi$ is a set of hypernetwork parameters.
The hypernetwork module generates the segmentation network parameters $\tilde{\mathbf\uptheta}$, such that the segmentation network parameters become a function of, and completely dependent on, \textit{(1)} the input hyperparameter values $\mathbf\upalpha$ and \textit{(2)} the hypernetwork parameters $\mathbf\upphi$. By re-parameterising the Combiner network in Eq.~\eqref{eq:network_shared}, the proposed segmentation network, equipped with the hypernetwork module, is referred to as the ``HyperCombiner network'':
\begin{equation}
  \mathbf{Y}^{\tau}=f(\mathbf{X}^{\tau}; h(\tilde{\mathbf\upalpha};\mathbf\upphi))
\end{equation}
where $\mathbf\upphi$ are the only trainable parameters, i.e.,~being updated by the gradients during the iterative backpropagation-based optimisation.

The linear mixture model in Eq.~\eqref{eq:linear} with parameters $\mathbf\upalpha$ is used here as an example and the combined class probabilities are
\begin{equation}
  Z=g^{lin}(\mathcal{Y};\mathbf\upalpha)=\sum_{\tau=1}^{\mathbf\uptau}\alpha_{\tau} \cdot \mathbf{Y}^{\tau}=\sum_{\tau=1}^{\mathbf\uptau}\alpha_{\tau} \cdot f(\mathbf{X}^{\tau};h(\tilde{\mathbf\upalpha}; \mathbf\upphi)),
\end{equation}
while the hyperparameter for the nonlinear stacking model (Eq.~\eqref{eq:nonlinear}) with the hypernetwork predicting $\tilde{\mathbf\upbeta}$ is omitted here for brevity. \ref{fig:net} illustrates an overall schematic for this network.

With the same training data set described as in Sec.\ref{sec:method.single-net} and Sec.\ref{sec:method.combine-net}, training the HyperCombiner with a hypernetwork added thus becomes optimising the hypernetwork parameters $\mathbf\upphi$:
\begin{equation}
\label{eq:hyper-opt}
\hat{\mathbf\upphi} = arg\min_{\mathbf\upphi} \sum_{j=1}^{J}\mathcal{L}(\mathbf{z}_{j}, \mathbf{t}_{j})
\end{equation}

The hyperparameters are randomly sampled from ranges of application interest during training of the HyperCombiner.

\subsection{Hyperparameter Estimation}
\label{sec:method.hp-estimation}
The HyperCombiner networks require hyperparameter values during inference. For the prostate mpMR imaging application, two estimation methods are described in this section, one approximating existing rules, such as those derived from PI-RADS, described in the following \ref{sec:method.hp-estimation.PI-RADS}, the other searching for the optimal ones among all plausible candidate rules, described in \ref{sec:method.hp-estimation.search}.

\subsubsection{Encoding { {PI}}-{ {RADS}} rules with condition and decision}
 \label{sec:method.hp-estimation.PI-RADS}
As discussed in \ref{sec:introduction}, radiological guideline scores individual image modalities, then combine these scores with a predefined rule. 

For the purpose of this study, binary classification decisions, clinically significant or not, are determined by a set of rules on the three available image modalities, T2W, DWI$_{hb}$ and ADC, based on PI-RADS. Such an example of the transition zone is provided in 
Table \ref{tab:condition_vectors}. Further discussion on the PI-RADS-derived rules for the transition- and peripheral lesions is provided in \ref{sec:app.alg}. In the experiments (\ref{sec:experiment}), separate hyperparameters $\hat{\mathbf\upalpha}^{(TZ)}$/$\hat{\mathbf\upbeta}^{(TZ)}$ and $\hat{\mathbf\upalpha}^{(PZ)}$/$\hat{\mathbf\upbeta}^{(PZ)}$, are specified for the lesions found in transition and peripheral zones, respectively, with their associated decisions $d_k^{(TZ)}$ and $d_k^{(PZ)}$. The estimation methods are summarised in this section as follows.

\begin{table}
  \caption{Example condition vectors and decisions for transition zone(TZ) lesions, $\tau$ $=$ 1,2 and 3 represent T2W, DWI$_{hb}$ and ADC, respectively. TZ denotes transition zone.} 
  \label{tab:condition_vectors}
  \begin{tabular}{lllll}
  \hline
    $\mathcal{C}$ & $p(\mathcal{C}\mid\tau=1)$ & $p(\mathcal{C}\mid\tau=2)$ & $p(\mathcal{C}\mid\tau=3)$ & $p(\mathcal{C})$ \\
    \hline
     TZ & Negative & Negative & Negative & Negative \\ 
     TZ & Negative & Negative & Positive & Negative \\ 
     TZ & Negative & Positive & Negative & Negative \\ 
     TZ & Negative & Positive & Positive & Positive \\
     TZ & Positive & Negative & Negative & Positive \\ 
     TZ & Positive & Negative & Positive & Positive \\ 
     TZ & Positive & Positive & Negative & Positive \\ 
    TZ & Positive & Positive & Positive & Positive \\
    \hline
\end{tabular}
\end{table}
\paragraph*{Condition vectors and associated decisions}
 
The PI-RADS rules can thus be encoded, with examples given in 
Table \ref{tab:condition_vectors}. Each row of $P(C\mid \tau)$ represents a ``condition vector'' $\mathbf{r}_k=[r_k^{\tau=1},r_k^{\tau=2},r_k^{\tau=3}]^\top, k=1, \ldots ,K$ and $r_k^{\tau}=\{0,1\}$ (here, $r_k^{\tau}=0$ and $r_k^{\tau}=1$ represent the modality $\tau$ showing negative and positive lesions, respectively). There are a maximum of $K=2^3$ unique combinations of possible deterministic conditions $\mathbf{r}_k$, and each of them has one associated ``decision'' $d_k=\{0,1\}$ (here, $d_k=0$ indicates that the final decision of the combined three modalities for the suspected lesion is negative, and $d_k=1$ indicating positive) to indicate the joint binary probability for all image modalities. That is,
\begin{equation}
\begin{aligned}
 \mathbf{R} =
 \left[\begin{array}{@{}cccccccc@{}}   0 & 0 & 0 & 0 & 1 & 1 & 1 & 1\\
   0 & 0 & 1 & 1 & 0 & 0 & 1 & 1\\
   0 & 1 & 0 & 1 & 0 & 1 & 0 & 1
\end{array}
 \right], 
 \mathbf{d} = 
 \left[\begin{array}{@{}c@{}}   0\\ 0\\ 0\\ 1\\ 1\\ 1\\ 1\\ 1
\end{array}
 \right]
\end{aligned}
 \label{eq:condition-vect}
\end{equation}
where $\mathbf{R}=[\mathbf{r}_1, \ldots ,\mathbf{r}_K]$ and $\mathbf{d}=[d_1, \ldots ,d_K]^\top$.

\paragraph*{Estimation with linear decisions}

To estimate the hyperparameters $\alpha_\tau$ with respect to the linear mixture model (\ref{sec:method.combine-net.linear}) in Eq.~\eqref{eq:linear},
substitute random variables $Y_i^\tau$ and $Z_i$, with $r_k^\tau$ and $d_k$, respectively. The resulting linear system in matrix form is:
\begin{equation}
  \mathbf{R}^\top \cdot \mathbf\upalpha = \mathbf{d}
\end{equation}
which can be solved by linear least-squares methods:
\begin{equation}
\label{eq:encode.ls}
  \hat{\mathbf\upalpha} = \frac {(\mathbf{R}\mathbf{R}^\top)^{-1} \mathbf{R} \mathbf{d}} {\| (\mathbf{R}\mathbf{R}^\top)^{-1} \mathbf{R} \mathbf{d}\| _1}
\end{equation}
This is equivalent to a \textit{linear classification} problem, with the denominator to ensure that the mixing hyperparameters sum to one $\sum^{\tau}_{\tau=1}\hat{\alpha_{\tau}}=1$.

\paragraph*{Estimation with nonlinear decisions}

The nonlinear stacking model (\ref{sec:method.combine-net.nonlinear}) can also be used as an equivalent \textit{logistic regression} problem. Substituting $\mathbf{R}$ and $\mathbf{d}$ into Eq.~\eqref{eq:nonlinear}, we have:
\begin{equation}
  \mathbf\upsigma([\mathbf{R}^\top,\mathbf{1}] \cdot \mathbf\upbeta) = \mathbf{d}
  \label{eq:encode.nonls}
\end{equation}
where $\mathbf{1}$ is vector of $K$ ones and $\mathbf\upsigma(\mathbf{a})=[\sigma(a_{0}), \ldots ,\sigma(a_{3})]^\top$ is the element-wise sigmoid function for a vector $\mathbf{a}=[a_{0}, \ldots ,a_{3}]^\top$. The estimated hyperparameter $\hat{\mathbf\upbeta}$ can be obtained by iterative methods minimising the mean-square residuals:
\begin{equation}
\hat{\mathbf\upbeta} = \min_{\mathbf\upbeta} \| d-\mathbf\upsigma([\mathbf{R}^\top,\mathbf{1}] \cdot \mathbf\upbeta)\| ^2_2.
\end{equation}

Both the linear and nonlinear hyperparameter estimation methods minimise the \textit{a prior} risk in making wrong decisions from multiple image modalities, as opposed to the empirical risk learned from data, which is discussed in the next section.

\subsubsection{Rule discovery as hyperparameter search}

\label{sec:method.hp-estimation.search}
Using the encoded condition vectors in 
Table \ref{tab:condition_vectors} may not be optimal, especially when the labels are not based on radiologists following these rules, such as Likert
scores~\citep{villers2006dynamic,jung2004prostate}, or not rule-based at all, such as histopathological labels. In such cases, optimising the rules is equivalent to optimising the hyperparameters. Automated hyperparameter tuning has been an
active research area for deep neural
networks~\citep{hoopes2021hypermorph,brock2017smash,lorraine2018stochastic},
and also a subject of the broad meta-learning
study~\citep{mantovani2015tune,bui2020optimal}. 

Using the proposed HyperCombiner, however, exhaustive search methods such as grid search become feasible, compared with the need to re-train the models for every sampled hyperparameter value. The trained HyperCombiner networks can be conditioned with different values sampled from each hyperparameter dimension of either the linear mixture models or the nonlinear stacking models during inference. In the following experiments and results, we also show the potential of discovering new rules that may perform comparably or even better than those directly derived from PI-RADS using grid search strategy.

\paragraph*{Sampling for the linear mixture model}
\label{sec:sampling_lin}
The linear mixture model takes $ \mathbf\upalpha$ as hyperparameters to combine the modality-specific outputs as a final output. Here, in the training process, hyperparameters ${\mathbf\upalpha}$ are sampled from a Dirichlet distribution of $\mathbf\uptau=3$ categories, which satisfies both constraints, $\alpha_{\tau} \in [0,1]$ and $\sum_{\tau=1}^{\mathbf\uptau}\alpha_{\tau}=1$, defined in \ref{sec:method.combine-net.linear}. During inference, samples are obtained with an equidistant interval of 0.1 in each hyperparameter dimension for a grid search.

\paragraph*{Sampling for the nonlinear stacking model}
 
Unlike linear mixture model, whose hyperparameters can be ``reasonably'' sampled from Dirichlet's space, the range and the sampling intervals of the hyperparameters of the non-linear stacking model are practically difficult to decide, or inefficient to sample equidistantly because of the nonlinear logistic functions involved. To investigate hyperparameters ${\mathbf\upbeta}$ that correspond to a set of interpretable and practically separable rules, different combinations of decisions \textbf{d} are sampled, based on solving the Eq.~\eqref{eq:encode.nonls} using a logistic regression model (a one-layer network). The regression fitting error, $\| d-\mathbf\upsigma([\mathbf{R}^\top,\mathbf{1}] \cdot \mathbf\upbeta)\| ^2_2$ as in \ref{sec:method.hp-estimation.PI-RADS} was used to reject the implausible combinations. Details of an acceptance--rejection algorithm can be referred to in \ref{sec:app.alg}. 

\subsubsection{Interpretation of the estimated hyperparameters}
 
\label{sec:method.hp-interp}
Here, we describe three ways to understand and then explain the estimated hyperparameter values and, in turn, discuss the potential clinical applications of these interpretations.

\paragraph*{Quantifying necessity for selective image acquisition}
 
Several different image modalities have inter-dependence in their acquisition. For example, computing ADC may require both DWI$_{hb}$ and DWI with low b values (DWI$_{lb}$), but the DWI$_{hb}$ and DWI$_{lb}$ may be acquired independently without each other. These specialised imaging sequencing with practical constraints and variable feasibility are discussed further in \ref{sec:results} for specific clinical context. Setting one or more hyperparameters to zero, in \eqref{eq:linear} and \eqref{eq:nonlinear}, allows the associated image modalities to be omitted during inference. Therefore, their respective added diagnostic values can be quantified by the difference in performance to the optimal values, using all available image modalities. When one or more image modalities are completely removed from prediction, the performance loss is quantified to investigate different data availability scenarios. It is essential to highlight that such necessity is quantified with respect to the use of machine learning models, which may differ from current clinical practice where radiologists yet have access to such tools.

\paragraph*{Quantifying diagnostic importance of image modalities}

Linear decisions (Eq.~\eqref{eq:linear}) is a weighted sum of predictions from
individual image modalities. Non-zero hyperparameter values $\mathbf\upalpha$ can be
used to compute the importance for each image modality, for example, using the
t-statistics~\citep{molnar2020interpretable}:
\begin{equation}
  T_{\hat{\alpha_{\tau}}} = \frac{\hat{\alpha}_{\tau}-\alpha_{\tau}^0}{\mathbf{SE}(\hat{\alpha_{\tau}})}
\end{equation}
where $\mathbf{SE}(\hat{\alpha_{\tau}})=\sqrt{C_{\tau\tau}}$ is the standard error representing the variance of the individual linear regression coefficients, where $C=\sigma^2_{\mathbf{d}}(\mathbf{R}\mathbf{R}^\top)^{-1}$ is the variance--covariance matrix and $\sigma^2_{\mathbf{d}}$ is the decision variance; and, $\alpha_{\tau}^0=0$ can be used to test the significance of image modalities, with other notations defined in \ref{sec:method.hp-estimation}. To quantify the relative importance between the ${\mathbf\uptau}$ predictions, $\alpha_{\tau}^0=\frac{1}{\mathbf\uptau}$ can be used to represent equal contributions. This $T_{\hat{\alpha_{\tau}}}$ directly measures how likely the combined decision will be changed due to the image modality $\tau$. 

Using the nonlinear decision rules (Eq.~\eqref{eq:nonlinear}), non-zero hyperparameter value $\mathbf\upbeta$ measures the log-ratio between the odds of the unit-incremented feature $Y^{\tau}_{i}+1$ and the estimated image modality-specific $Y^{\tau}_{i}$. Here, the odds of a specific image modality is positive against negative probabilities $odds(Y^{\tau}_{i})={p(\mathcal{C}=1\mid\tau)}:{p(\mathcal{C}=0\mid\tau)}=\frac{p(\mathcal{C}=1\mid\tau)}{1-p(\mathcal{C}=1\mid\tau)}$, where $Y^{\tau}_{i}={p(\mathcal{C}\mid\tau)}$. In other words, the exponential of the estimated hyperparameter equals the ratio of odds:
\begin{equation}
\mathit{e}^{\hat{\beta}_{\tau}}=\frac{odds(Y^{\tau}_{i}+1)}{odds(Y^{\tau}_{i})}
\end{equation}
which indicates how much the image modality-specific influence is on the prediction. 

In addition to the above general statistical interpretation, these estimated hyperparameters can be altered to assess their impact on metrics that are relevant to specific clinical applications, such as Type 1 and Type 2 errors for a downstream treatment decision following the machine learning-aided diagnosis or localisation. In this work, such sensitivity analysis was reported using the lesion-level recall and precision, designed for a class of tumour-targeting applications that use the MR-derived lesion locations, described in \ref{sec:exp.metrics}.

\paragraph*{Quantifying uncertainty due to varying decision}
 
The proposed HyperCombiner networks enable efficient comparison between different decision rules during inference. In addition to the optimum rule discovery described in \ref{sec:method.hp-estimation.search}, Monte-Carlo methods become feasible to estimate the variance of tumour localisation due to changing rules. This provides a means to quantify the type of uncertainty when image modality-related weightings cannot be estimated or applied precisely, a common case in using and reading the clinical MR images of prostate cancer.

\section{Experiment}

\label{sec:experiment}

\subsection{Data sets}

\label{sec:exp.data set}
 {The mpMR images were from several clinical trials
conducted at University College London Hospital, with a mixture of clinical
scanners, including a mixed biopsy and therapy patients. Further details of the
datasets, such as vendors and imaging protocols adopted in individual studies,
can be found in the original study papers and their published supplementary
materials, SmartTarget~\citep{hamid_smarttarget},
PICTURE~\citep{simmons_picture}, ProRAFT~\citep{orczyk_proraft},
Index~\citep{dickinson_index} and PROMIS~\citep{bosaily_promis}.}
All trial patients gave written consent, and the ethics were approved as part
of the respective trial protocols~\citep{hamid_smarttarget}. Part of the
dataset used in this study is publicly available via.\footnote{PROMIS Study
Dataset - Open Access Request. Retrieved from:
\url{https://ncita.org.uk/promis-data-set-open-access-request/}.}
Radiologist contours were obtained for all lesions with
Likert-scores $\geq$3 and served as ground-truth labels in this study. 22,
192, 325, 232 and 106 studies have 0, 1, 2, 3 and $\geq$4 lesions,
respectively. There are 1962 lesions in total, out of which 207, 1443 and 312
lesions are located in peripheral zone, transition zone and across peripheral
zone and transition zone.
The original in-plane dimension for T2W images ranges from 180~$\times~$180 to 640~$\times~$640 at a resolution of 1.31~$\times~$ 1.31~$\mathrm{mm}^{2}$ to 0.29~$\times~$ 0.29~$\mathrm{mm}^{2}$, respectively, and the thickness of T2W images ranges from 0.82 to 1~mm. The original in-plane dimension for DWI images ranges from 96~$\times~$94 to 456~$\times~$320 at a resolution of 3.41~$\times~$3.41~$\mathrm{mm}^{2}$ to $0.75\times0.75~\mathrm{mm}^{2}$, respectively, and the thickness of DWI images ranges from 3 to 5~mm.

All the image modalities were resampled to isotropic-sized voxels $1\times 1 \times 1 ~\text{mm}^3$ with a linear interpolator and applied linear intensity normalisation per modality ranging from 0 to 1. 
 {In this study, sequence alignment was performed using an image-to-scanner transformation, with T2-weighted (T2W) images serving as the reference coordinate system, as all the radiological lesions were annotated on T2W coordinate. No additional registration methods were employed to correct potential misalignment caused by factors such as organ deformation, patient movement, or imaging motion artefacts. These types of misalignment were not deemed severe in our data. Moreover, the potential benefits of registration for segmentation accuracy in the presence of misalignment to the reference image remain an open research question.}

 {For the experimental data split, 651 patients out of 850 patients had all three types of images, while a subset of patients had multiple studies and different high-b-valued diffusion images (e.g.~multiple b-value sequences that are all greater than 1400). As a result, the total number of cases used is 751. To ensure no data leakage, these cases are randomly divided on a patient-level into 500 training, 124 validation, and 127 test cases, such that data from the same patient will not be grouped in different sets.}

\subsection{Implementation}

\label{hyperimple}
All the experiments using UNet were implemented with TensorFlow 2.4, and nnUNet-based experiments were implemented with Pytorch 2.0. All the networks were trained on an NVIDIA Tesla V100 GPU with 32~GB memory.

\subsubsection{Base networks}

The base networks are built on the widely used UNet~\citep{ronneberger2015u},
which consists of an encoder--decoder with skip connections. The input to the
base network was a batch of individual type images of size $N\times96\times96\times96\times1$, where
N is the batch size. A 3D convolutional layer was followed by Leaky Relu
activation as the basic block in the encoder and decoder. The basic blocks had
16, 32, 32 and 32 channels in the encoder and 32, 32, 32, 32 and 16 channels in
the decoder, where skip connections were used to build an information bridge
between the encoder and decoder. We applied max-pooling with a factor of 2
after each basic block to reduce the spatial dimension four times in the
encoder and then upsampled the feature maps of the basic blocks in the decoder
until they reached the original input dimension. The last layer of the base
network was a $1~\times~1~\times~1$ convolutional layer followed by a sigmoid activation
that generates a probability map representing the segmentation mask.

These three networks were separately trained with loss function Eq.~\eqref{eq:loss} for 100 epochs using the ADAM optimizer with a learning rate of $1\times10^{-4}$. 

\subsubsection{Multi-channel baseline networks}
 
\label{sec:exp.baseline}
 {
We implemented UNet and nnUNet as baseline segmentation networks. These
networks were designed to take concatenated three image modalities as inputs
and were used for comparison with the Combiner networks. Implementation of UNet
was similar to the Base network but with three input channels. No data
augmentation was performed with the UNet implementation, providing a reference
performance in this study. For nnUNet experiments, the implementation described
in~\citet{isensee2021nnu} was utilised. The built-in data augmentation in the
nnUNet was used alongside other segmentation-performance-improving techniques,
such as deep supervision, connected component analysis and mirror augmentation
ensembles.} 

These networks were rained with loss function Eq.~\eqref{eq:loss} for 100 epochs using the ADAM optimizer with a learning rate of 1$\times10^{-4}$. 

\subsubsection{Combiner networks}

\label{sec:exp.combiner}
The Combiner networks share the same architecture as the baseline approaches, namely UNet and nnUNet networks. However, in the case of the Combiner networks, rather than using concatenation as input, each modality is taken as input separately, enabling them to contribute independently to the decision-making process in subsequent stages. The network weights are shared among the three modalities. Additionally, we incorporated another comparison Combiner, where the base networks have no shared weights. An auxiliary combining module was implemented to combine the modality-independent output from three base networks. In this application, we used the fixed hyperparameters $\mathbf{\alpha}$ or $\mathbf{\beta}$ for training and testing, and the derived hyperparameters are stated in \ref{Res:condition_vectors}.

The network was trained with loss Eq.~\eqref{eq:loss} using ADAM optimizer, with a learning rate equal to $1\times10^{-5}$.

\subsubsection{ {HyperCombiner} networks}

The HyperCombiner network has the same structure as the Combiner network,
implicitly representing the convolutional kernel as a function of
hyperparameters by replacing the convolutional layer in Combiner with a
hyperparameterized convolutional layer~\citep{hoopes2021hypermorph}. The
weights (and biases) of the Combiner network were generated by an auxiliary
hypernetwork. The combination of this hypernetwork and the Combiner network
itself is referred to as the HyperCombiner network, as shown in Fig.
~\eqref{fig:net} (right). The hypernetwork consisted of 4 fully-connected layers,
each with 64 units and followed by ReLU activation except for the final layer,
which used Tanh activation. The hypernetwork provided all the weights and
biases required in the Combiner network; therefore, only parameters in the
hypernetwork were trainable. 

\paragraph*{Train with linear mixture models}
\label{sec:imp.linear}
The Combiner network of linear mixture model in HyperCombiner was the same as that of Combiner networks. During training, the hyperparameters were randomly sampled as described in \ref{sec:sampling_lin}. All HyperCombiner networks with linear mixture models were trained with the loss function in Eq.~\eqref{eq:loss} for 400 epochs using the ADAM optimizer, with a learning rate of 1$\times10^{-6}$.

\paragraph*{Train with nonlinear stacking models}
\label{sec:imp.nonlinear}

The sampling was required from each iteration of training the HyperCombiner networks. To avoid cumbersome data transfer between the sampling and neural network training, Algorithm \ref{alg:nonlinear}{1} in \ref{sec:app.alg} was run in advance for sufficient iterations and these pre-computed hyperparameter values in neural network training. Each network was trained using the ADAM optimizer with a 1$\times10^{-6}$ learning rate for 400 epochs.

\subsection{Evaluation}
\label{sec:exp.metrics}
All aforementioned experiments were tested with both voxel-level and lesion-level evaluation metrics. {During evaluation, we used a threshold value of 0.5 to generate a binary segmentation mask directly from the softmax output and excluded only extremely small regions smaller than 27 pixels, without additional post-processing.}

\subsubsection{Voxel-level metrics}

DSC and Hausdorff Distance (HD) are used as evaluation metrics at voxel-level. DSC measures the overlap between the predicted segmentation $Y_p$ and the ground-truth $Y_g$, $\mathcal DSC =2~\times~\mid{Y_p\cap{Y_g}}\mid/(\mid Y_p\mid+\mid Y_g\mid)$. HD measures the greatest surface distance between the boundaries of the predicted segmentation and the ground-truth.
 We report the 95th percentile of surface distances as a robust alternative, denoted as $D_{HD}$. 

\subsubsection{Lesion-level metrics}

The lesion-level evaluation metrics used in this study are adapted from those
found in the object detection literature~\citep{electronics10030279}, for our
intended clinical targeting application, such as progression monitoring,
targeted biopsy and focal therapy. Since multiple prostate lesions are often
present in a single prostate MR image, directly applying object detection
Metrics have been found to be challenging in the application of multifocal cancers. In
this study, we used modified asymmetric lesion-level evaluation metrics to
evaluate the multifocal segmentation output, and further discussion of the
motivation for these metrics is detailed in~\citet{yan2022impact}. 

For each of $N$ ground-truth lesions $\{Y_g^n\}_{n=1,...,N}$, it is considered as a true-positive lesion if it has overlap with any of the $M$ predictions \{$Y_p^m\}_{m=1,...,M}$, single or multiple, that is greater than a pre-defined overlap threshold $s^{GT}$, otherwise false-negative. Thus, \(\mathcal{S}^{GT}=\sum_{m=1}^M (Y_p^m\cap{Y_g^n})/{Y_g^n}\), with the superscripts $GT$ indicating the ground-truth-based definitions, with which false-positive lesions is not defined. The recall thus can be computed as
\begin{equation}
recall^{GT}=TP^{GT}/(TP^{GT}+FN^{GT}),
\end{equation}
where $TP^{GT}$ and $FN^{GT}$ are the numbers of true-positive and false-negative lesions using the ground-truth-based definitions, respectively.

For individual predicted lesions $Y_p^m$, a true-positive lesion requires the overlap with ground-truth regions $Y_g^n$ to be greater than $s^{Pred}$, otherwise false-positive. Thus, $\mathcal{S}^{Pred}=\sum_{n=1}^N (Y_g^n\cap{Y_p^m})/{Y_p^m}$, with the superscripts $Pred$ for the prediction-based definitions and undefined $FN^{Pred}$. Therefore,
\begin{equation}
precision^{Pred}=TP^{Pred}/(TP^{Pred}+FP^{Pred}).
\end{equation}

\subsection{Comparison between Different Networks}

We describe as follows a set of comparison experiments to demonstrate that (a) the proposed HyperCombiner networks are capable of cancer segmentation, compared with the baseline networks, by testing several scenarios described as follows, and (b) hyperparameters can be conditioned accurately on the proposed HyperCombiner networks, compared with the individually trained Combiner networks.

\subsubsection{Using all three image modalities}

Baseline performance of cancer segmentation is established using all image modalities available. The quantified performance between the baseline network (\ref{sec:exp.baseline}), Combiner networks and HyperCombiner, are compared when they are trained using all three available image modalities. 

Four Combiner networks based on linear mixture models for two zonal lesions, each being trained with different decision values that represent (1) equally-weighted hyperparameters and (2) PI-RADS-derived condition vectors. Two Combiner networks based on the nonlinear stacking models are trained with the PI-RADS-derived condition vectors. Two further HyperCombiner networks are trained and compared for the linear mixture- and the nonlinear stacking models, while the same set of decisions can be instantiated and tested during inference.

\subsubsection{Using individual image modalities}

Combiner networks and HyperCombiner networks, when only individual image modalities are available, are compared with the prediction accuracy of the Base network with a single image modality. Here, the Combiner networks with single image modality input are considered baseline performance. These experiments provide a predictive reference performance concerning each image modality. 

\subsubsection{Varying hyperparameter values}

\label{sec:exp.compare-hp}
With two or more types of images available, the difference in performance was also tested between the Combiner networks and HyperCombiner networks, with varying hyperparameter values. Among possible permutations, setting individual hyperparameters to zero indicates predicting without using certain types of images. As described in the above section, predicting using individual image modalities is a special case when two of the hyperparameters are zero.

The permutation was possible during inference for HyperCombiner, while the Combiner models need to be trained separately for each set of hyperparameter values. The following were compared between the two networks: (1) PI-RADS-derived hyperparameters, described in \ref{sec:method.hp-estimation.PI-RADS}, (2) the sampled hyperparameter values (including those being zeros), described in \ref{sec:method.hp-estimation.search}, and (3) for the linear mixture model-based networks, equal hyperparameters between all image modalities.

\subsection{Hyperparameter Analysis}

\label{sec:exp.hyperanalysis}
This section describes experiments from the HyperCombiner networks, taking advantage of efficient conditioning for experimental purposes. 
 Network performance using the metrics described in \ref{sec:exp.metrics} was used for comparing different rules described below and, when applicable, statistics and a possible explanation of the hyperparameter values described in \ref{sec:method.hp-interp}, are discussed. 

\subsubsection{Comparing different rules}
 
Comparing different rules is equivalent to comparing HyperCombiners inferred with different hyperparameter values, including PI-RADS derived rules, random sampling rules, and specially assigned hyperparameter values, such as equal weighting for all image modalities.

\subsubsection{Rule discovery}
 
The search for the optimum rule-defining hyperparameters was demonstrated with a grid search. The sampling of these parameters is described in \ref{sec:method.hp-estimation.search}. For the HyperCombiner with linear mixture models, $\mathbf\upalpha$ was sampled between $[0,1]$ with equidistant intervals of 0.1. For the HyperCombiner with a nonlinear stacking model, $\mathbf\upbeta$ was first randomly sampled before being sorted for each hyperparameter for the grid search.
Each possible combination of the hyperparameters ${\alpha_{\tau}}$ and ${\beta_{\tau}}$ was then tested with the evaluation metrics.

\section{Results}

\label{sec:results}

\subsection{{ {PI}}-{ {RADS}}-derived Decisions}

\label{Res:condition_vectors}
Given a constant \textbf{R} that contains all possible deterministic condition vectors, as in Eq.~\eqref{eq:condition-vect}, $\mathbf{d}^{(TZ)}$ and $\mathbf{d}^{(PZ)}$ denote decisions for identifying lesions found in transition and peripheral zones, respectively. For comparison, $\mathbf{d}^{(WG)}$ is also considered for lesions found in the whole gland regardless of their zonal location. The PI-RADS derived hyperparameter values are reported in Table \ref{tab:results_pirads_fit}, together with residuals and importance-representing statistics. \ref{sec:app.rules} contains the details in obtaining these PI-RADS-derived decisions and the fitted hyperparameters.

\begin{table*}[htb]
    \centering
    \caption{PI-RADS-derived decisions and the fitted hyperparameters, together with the corresponding residuals and importance-representing statistics (Sec.~\ref{sec:method.hp-interp}). The hyperparameters correspond to the three types of images in the same order of T2W, DWI$_{hb}$ and ADC, following the optional bias $\hat{\beta}_0$ in the nonlinear models\protect\footnotemark{}, for all the presented results. $No.^{rule}$ is the decimal value of decision vector $\textbf{d}$.}
    \label{tab:results_pirads_fit}
    \resizebox{\textwidth}{!}{
    \begin{tabular}{ccccccc}
    \hline
     Zones & $No.^{rule}$ & Decision $\mathbf{d}^\top$ & combining & Hyperparameters $\mathbf\upalpha(\mathbf\upbeta)$&  Residuals & Quantified diagnostic importance\\
     \hline
         \multirow{2}{*}{WG} & \multirow{2}{*}{63} & \multirow{2}{*}{[0 0 1 1 1 1 1 1]} & linear & $\hat{\mathbf\upalpha}^{(WG)}=[0.45,0.45,0.10]^\top$ & 0.0781 &$T_{\hat{\mathbf\upalpha}}= [2.2048, 2.2048, 0.4410]$\\
    \cline{4-7}
    &&&nonlinear &  $\hat{\mathbf\upbeta}^{(WG)}=[18.17, 18.17, -0.20, -8.53]^\top$ & $1.35\times 10^{-7}$ & $e^{\hat{\mathbf\upbeta}}$=$[ 7.78\times 10^7,7.78\times 10^7,   9.80\times10^{-1},   1.97\times10^{-4}]$ \\ \hline
     \multirow{2}{*}{TZ} & \multirow{2}{*}{31}&\multirow{2}{*}{[0 0 0 1 1 1 1 1]} & linear & $\hat{\mathbf\upalpha}^{(TZ)} = [0.60,0.20,0.20]^\top$ & 0.0625 & $T_{\hat{\mathbf\upalpha}}=[2.3664, 0.7888, 0.7888]$\\
     \cline{4-7}
     &&& nonlinear & $\hat{\mathbf\upbeta}^{(TZ)}=[30.67, 14.84, 14.84, -22.36]^\top$ & $1.77\times 10^{-7}$ & $e^{\hat{\mathbf\upbeta}}=[  2.09\times 10^{13},2.79\times 10^{6}, 2.79\times 10^{6}, 1.95\times 10^{-10}]$\\
     \hline
     \multirow{2}{*}{PZ} & \multirow{2}{*}{119}& \multirow{2}{*}{[0 1 1 1 0 1 1 1]} & linear & $\hat{\mathbf\upalpha}^{(PZ)}=[0.10,0.45,0.45]^\top$ & 0.0781 &$T_{\hat{\mathbf\upalpha}}=[2.3094, 2.3094, 2.3094]$\\
     \cline{4-7}
     &&& nonlinear & $\hat{\mathbf\upbeta}^{(PZ)}=[-0.20, 18.17, 18.17, -8.53]^\top$  & $1.18\times 10^{-6}$ & $e^{\hat{\mathbf\upbeta}}=[9.80\times10^{-1}, 7.78\times 10^7,7.78\times 10^7, 1.97\times10^{-4}]$\\
     \hline
    \end{tabular}}
    
\end{table*}
\footnotetext{$\mathbf\upbeta$ is listed in order of $[\beta_1, \beta_2, \beta_3, \beta_0]^{\top}$ in all tables for readability, where the first three $\beta$s are ``stacking parameters'' in Eq.\eqref{eq:encode.nonls} and $\beta_0$ is bias term.}

As indicated by Quantified Diagnosis Importance in
Table \ref{tab:results_pirads_fit}, TZ lesions are more indicative with the T2W, although they require, like the PZ lesions, all three types of T2W, DWI$_{hb}$ and ADC. This is consistent with the verbal description of the PI-RADS rules that determining positive lesions needs both T2W and DWI modalities. An interesting discussion between the requirement of either or both ADC and DWI$_{hb}$ can be facilitated with their necessity analysis (\ref{sec:method.hp-interp}), presented in \ref{sec:results.interp.availability}.

\subsection{Network Performance}

The results in terms of network performance metrics, both voxel-level and lesion-level, are presented in 
Table \ref{tab:overallresults}, with their comparisons summarised in this section as follows.

\begin{table*}
  \caption{The table provides a summary of the network performance. The first four rows display the results of the Base networks, followed by the majority voting results based on the Base networks. The next four rows (fifth to eighth) showcase the performance of the baseline network and Combiner with PI-RADS rule, utilising UNet as the backbone. Similarly, the subsequent four rows (ninth to twelfth) present the results using nnUNet as the backbone. The last four rows represent the outcomes of the HyperCombiner networks. The symbols $\hat{\mathbf\upalpha}$ and $\hat{\mathbf\upbeta}$ denote the fixed PI-RADS-derived hyperparameters for the linear and nonlinear models, respectively. On the other hand, the symbols $\mathbf\upalpha^{*}$ and $\mathbf\upbeta^{*}$ indicate the best rules obtained from the hyperparameter search, which may vary for WG, TZ, and PZ. For further details, please refer to Table \ref{tab:linear} and \ref{tab:gridsearch_voxel}. ``Backbone with $\times3$'' indicates that the Combiner trained with separate segmentation networks that do not share weights.}
  \label{tab:overallresults}
  
      \resizebox{\textwidth}{!}
    {
    \begin{tabular}{cc|cccc|cccc|cccc}
        \hline
         \multirow{2}{*}{Methods}& \multirow{2}{*}{backbone} & \multicolumn{4}{c|}{WG}&\multicolumn{4}{c|}{TZ} & \multicolumn{4}{c}{PZ}\\\cline{3-14}
             & & $DSC$ & $D_{HD}$ & $Rec.*^{GT}$ &$Prec.*^{Pd}$ &  $DSC$ & $D_{HD}$ & $Rec.*^{GT}$ &$Prec.*^{Pd}$ & $DSC$ & $D_{HD}$ &$Rec.*^{GT}$ &$Prec.*^{Pd}$\\
        \hline
        Base(T2W) & UNet  & 0.26(0.18)&19.55(9.65)&0.69&0.49  &0.09(0.13)&22.09(13.11)&0.38&0.24	&0.33(0.19)&17.88(10.33)&0.73&0.65\\
        Base(DWI$_{hb}$) & UNet  & 0.27(0.19) & 17.40(11.17)& 0.77 & 0.47  & 0.12(0.11)& 24.35(13.80)& 0.26 & 0.25 & 0.32(0.22) & 17.18(10.77) & 0.75 & 0.65 \\
        Base(ADC)& UNet   & 0.21(0.15)&17.75(9.18)&0.57&0.51  &0.10(0.14)&23.45(11.86) & 0.30&0.24&0.27(0.19)&18.37(10.33)&0.54&0.56 \\
        MajorityVoting & UNet & 0.28(0.20) & 16.52(8.60) & 0.72 & 0.56 & 0.10(0.14) &18.31(8.48) & 0.43 & 0.21 & 0.31(0.20) & 16.77(9.57) & 0.68 & 0.61 \\
        \hline
         Baseline& UNet & 0.31(0.18) & 17.43(10.21) & 0.72 & 0.61 &0.13(0.10) &21.77(12.28) &0.45 & 0.31& 0.35(0.21) &16.72(10.72) & 0.66 &0.66\\
        
        Combiner$^{(\hat{\mathbf\upalpha})}$ & UNet  & 0.29(0.17) & 17.54(8.46) & 0.68 &0.66 & 0.12(0.11) & 18.66(9.38) & 0.47 & 0.33 & 0.33(0.20) &{16.22}(8.95)& 0.73 & 0.69\\
         Combiner$^{(\hat{\mathbf\upalpha})}$ & UNet$\times3$  & 0.29(0.16) & 17.43(9.44) & 0.63 &0.66 & 0.13(0.11) & 18.72(10.01) & 0.48 & 0.35 & 0.33(0.20) &16.19(8.76)& 0.69 & 0.65\\

        Combiner$^{(\hat{\mathbf\upbeta})}$ & UNet  & 0.30(0.18) &	17.57(8.25) & 0.78 & 0.66 & 0.13(0.13) & 18.21(8.59) &0.64 & 0.35 & 0.33(0.20) &17.46(9.35) & 0.74 & 0.65\\ 

        \hline
         Baseline & {nnUNet} & 0.42(0.21) &16.19(12.63) & 0.73 & 0.67 & 0.19(0.20) & 17.32(11.96) &0.48 & 0.53 &0.45(0.21) &15.49(12.75) & 0.71 & 0.66 \\
        
        Combiner$^{(\hat{\mathbf\upalpha})}$ & {nnUNet}  & 0.40(0.20) & 17.63(12.34) & 0.68 & 0.72 & 0.17(0.20) &17.92(12.75) & 0.36 & 0.58 & 0.43(0.21) &16.29(12.56) & 0.65 & 0.71  \\
        Combiner$^{(\hat{\mathbf\upalpha})}$ & {nnUNet$\times3$}  & 0.40(0.20) & 17.17(12.44) & 0.72 & 0.66 & 0.18(0.21) &17.78(12.03) & 0.42 & 0.55 & 0.43(0.21) &16.24(12.72) & 0.69 & 0.67  \\ 
        Combiner$^{(\hat{\mathbf\upbeta})}$) &  {nnUNet} & 0.42(0.21) & 15.43(13.50) & 0.76 & 0.63 & 0.25(0.23) & 13.07(13.37)& 0.51 & 0.51 & 0.45(0.22) &13.98(13.55) &0.71 & 0.65  \\ 
        \hline
        HypComb$^{(\hat{\mathbf\upalpha})}$ & UNet &0.29(0.21)&17.71(9.62)&0.67&0.58 &0.13(0.11) &19.52(7.55)&0.44 &0.37& 0.30(0.20)&17.16(9.26)&0.68 &0.73\\ 
       
        HypComb$^{(\hat{\mathbf\upbeta})}$ & UNet & 0.31(0.20) & 17.84(9.39) & 0.79 & 0.46 &0.11(0.18)&19.43(8.25)&0.55 &0.26&0.33(0.19)&17.57(9.35)&0.67&0.64\\ 
        \hline
        HypComb$^{(\mathbf\upalpha^{\star})}$ & UNet & 0.33(0.22)& 16.74(9.81)& 0.47 &0.51  &0.16(0.08) &19.58(10.53) &0.48 &0.22 &0.36(0.24)&17.32(10.46) & 0.66&0.50\\
       
        HypComb$^{(\mathbf\upbeta^{\star})}$ & UNet & 0.33(0.20) & 17.40(9.65) & 0.81 & 0.50 & 0.16(0.15) &18.81(8.99) & 0.58 & 0.28 & 0.37(0.21)&16.90(9.23) & 0.75 & 0.60\\
        \hline
    \end{tabular}}   
   
\end{table*}

\subsubsection{The Base networks}

The base networks segmented cancers from each individual image modalities, with the $DSC$ results ranging between [0.10, 0.12] in TZ and ranging between [0.27, 0.32] in PZ and the $D_{HD}$ results that differed in their variances (i.e.~St.D. range [9.2, 11.2] in WG). DWI$_{hb}$ always 
gave the best results regardless of the zones in which lesions were located.
In terms of the lesion-level metrics, DWI$_{hb}$ yielded the highest sensitivity ($Recall{*}^{GT}$ $=$ 0.75) in PZ, while T2W had the highest sensitivity ($Recall{*}^{GT}$ $=$ 0.56) among all three modalities in TZ. T2W and DWI have comparable positive predictive value $Precision{*}^{pred}$ in both TZ and PZ.

\subsubsection{Multi-channel {UNet} baseline network}

The UNet baseline network achieved a $DSC$ of $0.31~\pm~0.18$ and a $D_{HD}$ of $17.43~\pm~10.21~\mathrm{mm}$ in the WG, together with its lesions-level accuracy for lesions found in different zones also provided in 
Table \ref{tab:overallresults}. 
 {The nnUNet baseline outperformed the UNet baseline, achieving superior segmentation results. Specifically, the DSC values for the nnUNet baseline were $0.42~\pm~0.21$, $0.19~\pm~0.20$ and $0.45~\pm~0.21$ in the WG, TZ, and PZ, respectively. By integrating nnUNet and applying predefined combining rules, we were able to achieve more accurate segmentation results, whilst showing consistent conclusions in sequence comparison and rule discovery applications regardless of what segmentation model was used.
}

 {The Majority Voting approach was also tested to combine modality-independent predictions, serving as a non-parametric late-fusion method for reference. Its results are summarised in 
Table \ref{tab:overallresults}, and interestingly, it had a lower HD than the baseline UNet, albeit a lack of significance (p $=$ 0.053), while maintaining a comparable DSC.}

These results serve as a benchmark for evaluating the performance of the widely used end-to-end segmentation networks. The accuracy at both voxel-level and lesion-level for lesions located in different regions of the gland all surpassed that of the Base networks. These findings provide evidence supporting the enhanced diagnostic value of mpMR images compared to individual image modalities, particularly when combined with machine learning assistance.

\subsubsection{The Combiner networks}
\label{sec:results-comb}
Overall, the Combiner networks trained with PI-RADS-derived hyperparameters achieved comparable results to the baseline networks in terms of voxel- and lesion-level evaluation results, as shown in 
Table \ref{tab:overallresults}. {As nnUNet consistently outperformed UNet as Base networks, this also led to an improved performance as nnUNet-based Combiners, achieving DSC ranging from 0.17 to 0.25 in TZ and 0.43 to 0.45 in PZ. However, it is interesting to note that despite the improved rule-modelling capability of Combiners, there was a slight decline in overall predictive performance observed in the linear mixture model compared to baseline methods, both with UNet and nnUNet-based Combiners.}
From the same table, for PZ lesions with UNet-based Combiner, the DSC decreased from 0.35 to 0.33 ($p$-value $=$ 0.02) compared to the baseline method. This indicates that the linear rules derived from PI-RADS may not be the most optimal in this case.
 {Additionally, we incorporated another comparison Combiner, where the base networks have no shared weights, indicated by ``backbone with $\times3$'' in 
Table \ref{tab:overallresults}. These experiments performed very similarly to the shared weights network.}

\begin{figure*}
\caption{The first and second row of figures compared the DSC between Combiner and HyperCombiner based on UNet backbone in terms of the linear mixture and nonlinear stacking models, respectively. Note that $x$ coordinate was the $No._{Rule}$ of randomly selected combinations of hyperparameters.\label{fig:valid_hyper}}
\includegraphics[width=\linewidth]{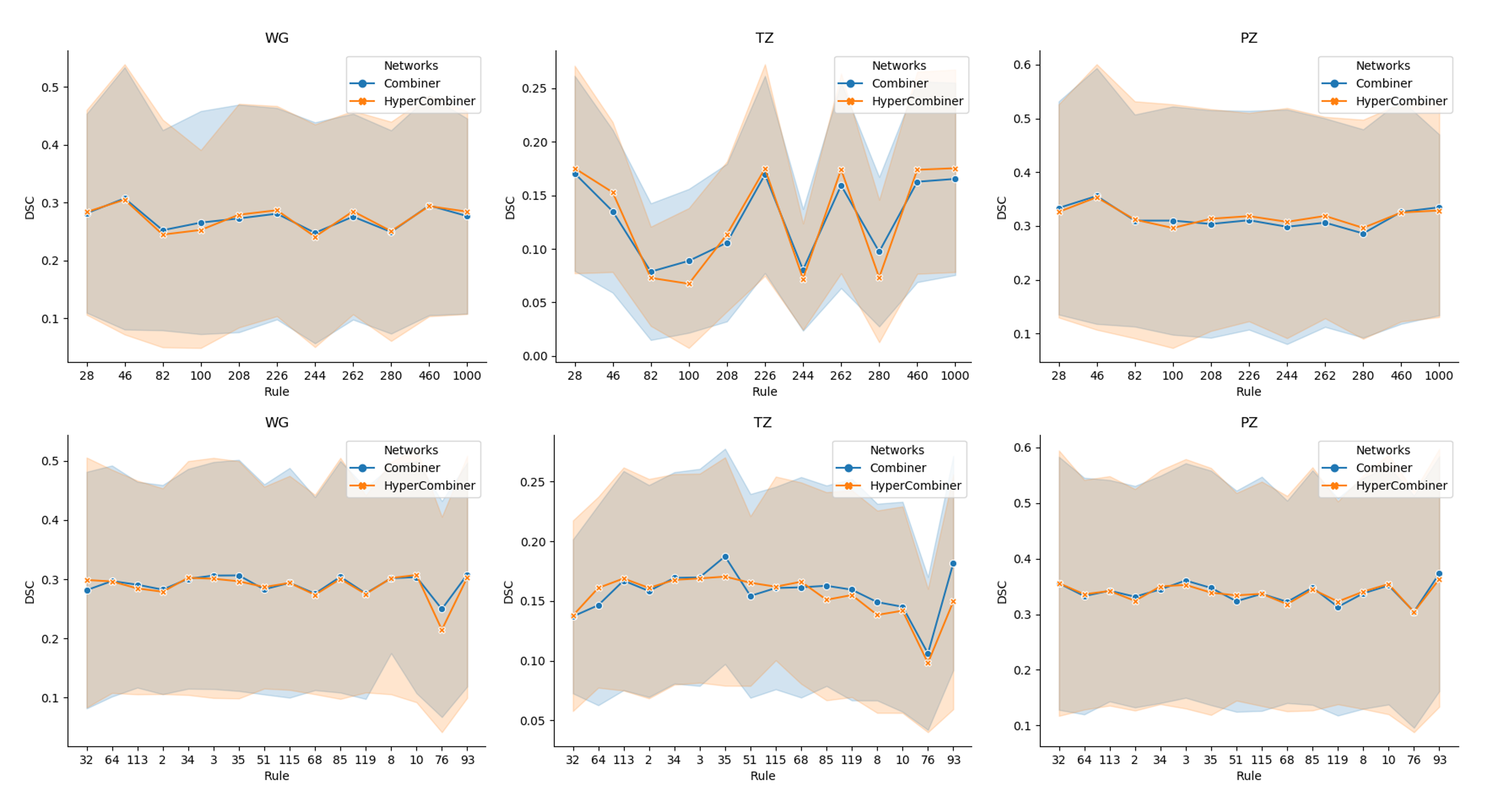}
\end{figure*}

\subsubsection{The {HyperCombiner} networks}

The validity of the HyperCombiner networks is bench-marked by the segmentation metrics based on the Combiner networks for permuted hyperparameter values and different lesion locations, as shown in \ref{fig:valid_hyper}. The randomly selected ten combinations of hyperparameters were used to demonstrate the DSC values between the HyperCombiner and Combiner networks at the same hyperparameter values. The results showed that, in general, the single HyperCombiner networks are capable of replacing different Combiner networks trained with varying hyperparameter values.

To examine the correlation between the DSC of individual lesions with sizes, we conducted a statistical analysis of ``big'' and ``small'' lesions regarding a threshold diameter of 15~mm. There were 372 big lesions and 252 small lesions in the test dataset, with corresponding lesion-level Dice scores of 0.26 and 0.10, respectively. This analysis revealed a statistically significant difference (Mann--Whitney U test, $p$-value of $1.49\times 10^{-17}$) between the two categories, suggesting that the large standard deviation observed in \ref{fig:valid_hyper} may be associated with the variation in lesion size.

\subsection{Hyperparameter analysis and interpretation}

\subsubsection{Image type availability of linear mixture model}

\label{sec:results.interp.availability}

Table \ref{tab:linear} shows how the availability of image modalities affected the DSC of the linear mixture model by setting the corresponding hyperparameters of unavailable types to zeros. DWI$_{hb}$ only and T2W only achieved the best performance in TZ and PZ, respectively. 
The Best results from PZ and TZ were both yielded by T2W and DWI$_{hb}$ combinations, albeit with different hyperparameters, suggesting that the combination of T2W and DWI$_{hb}$ may be adequate without ADC.

It may be intuitive that the inferior lesion localisation ability from ADC may adversely affect the overall performance with such hyperparameter-ised rules, while it may not be the case for an end-to-end representation learning such as the baseline network.

\begin{table*}
  \caption{The results of HyperCombiner based on UNet backbone with linear mixture model that was evaluated under various fixed hyperparameters, and the three types of images are in the same order of (T2W, DWI$_{hb}$, ADC) for all the presented results. The last three rows showed the best rules that were ranked by grid searching on the validation set and tested on test set. }
  \label{tab:linear}

      \resizebox{\textwidth}{!}{
    \begin{tabular}{ccccccccccccccc}
        \hline
        \multirow{2}{*}{Hyperparameters - $\mathbf\upalpha$}&
        \multicolumn{4}{c}{WG} & \multicolumn{4}{c}{TZ} & \multicolumn{4}{c}{PZ}\\\cline{3-15}
        &&DSC&$D_{HD}$&$Rec.^{GT}$&$Prec.^{pd}$&DSC&$D_{HD}$&$Rec.^{GT}$&$Prec.^{pd}$&DSC&$D_{HD}$&$Rec.^{GT}$&$Prec.^{pd}$&\\ \hline
       \multirow{3}{*}{Single Image Type}
        &[1,0,0]   & 0.30(0.21)  &18.19(10.13)&0.55&0.37  &0.12(0.12) &19.39(10.71)&0.34&0.10	&0.34(0.23) &17.76(10.25)&0.67&0.48\\
        
        &[0,1,0]   & 0.28(0.19)  &19.10(9.91)&0.62&0.33	&0.15(0.10) &19.53(9.85) &0.52&	0.18&0.33(0.22) &18.73(10.07) &0.73&0.45\\
        
        &[0,0,1]   & 0.23(0.18) &18.33(9.16)&0.35&0.29	&0.10(0.13) &20.12(10.05) &0.25&0.13	&0.27(0.20) &18.08(10.44) &0.50&0.41\\
        \hline
        \multirow{3}{*}{Two Image Types}
        &[0.5,0.5,0] &0.33(0.22)& 16.74(9.81)& 0.47 & 0.51 	&0.16(0.11)&19.20(10.43) &0.48&0.20 & 0.36(0.24)&17.32(10.46) & 0.66&0.50\\
        
        &[0,0.5,0.5] &0.28(0.20) &17.14(9.51) &0.63&0.42	&0.14(0.15) &19.22(10.04)&0.20&0.23	&0.31(0.22) &17.00(10.39) &0.71&0.49\\
        
        &[0.5,0,0.5] &0.28(0.22) &17.20(10.00)&0.64&0.36  & 0.10(0.15) &22.36(12.78)&0.24&0.23 &0.31(0.23) &17.62(11.01) &0.78&0.44\\
        \hline
        {Equal Weights}
        &[$\frac{1}{3}$,$\frac{1}{3}$,$\frac{1}{3}$] &0.30(0.20) &17.28(8.67)&0.68&0.44 &0.13(0.12) &17.55(7.74)&0.54& 0.19 &0.33(0.22) &16.84(9.67)& 0.63 &0.49\\
        \hline
        {Best for WG}
        &[0.5,0.5,0.0] &0.33(0.22)& 16.74(9.81)& 0.47 &0.51 &/ &/&/&/ &/&/&/&/\\
        {Best for TZ}
        &[0.4,0.6,0.0] &/ &/&/&/ &0.16(0.08) &19.58(10.53) &0.48 &0.22 &/&/&/&/\\
        {Best for PZ}
        &[0.5,0.5,0.0] &/ &/&/&/ &/ &/&/&/ &0.36(0.24)&17.32(10.46) & 0.66&0.50\\

       \hline
    \end{tabular} }

\end{table*}
\subsubsection{Comparison of decision rules}

\paragraph*{Results of linear decision rules}

\begin{figure*}
\caption{Grid search results on validation dataset of linear mixture learning heat map with various combinations of hyperparameters ($\alpha_1, \alpha_2, \alpha_3$), where $\alpha_1+\alpha_2+ \alpha_3=1$. Thus, we just show heat map of $\alpha_1$ and $\alpha_2$, and corresponding $\alpha_3= 1-\alpha_2- \alpha_1$.\label{fig:heat_linear}}
\includegraphics[width=\linewidth]{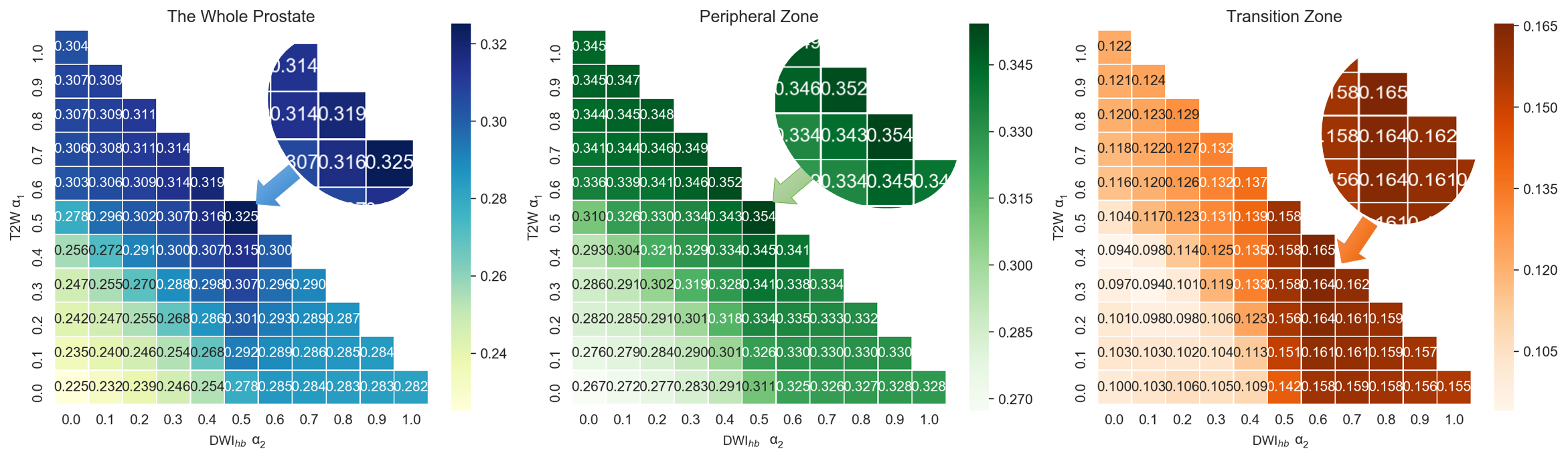}
\end{figure*}

 \ref{fig:heat_linear} shows DSC accuracy of 55 linear rules using HyperCombiner in the grid search on validation dataset. The greater DSC values are found in the upper left-, upper left- and lower right parts of the triangle for the WG, PZ and TZ lesions, respectively. The observation implies that lesions in PZ and TZ were more accurately localised with more significant contributions from T2W and DWI$_{hb}$, respectively, compared to the lower left part of the triangle that represents the alternatively higher contribution from ADC. To a lesser degree, T2W was found to be more important than DWI$_{hb}$ for PZ lesions and \textit{vice versa} for TZ lesions. However, decreasing the weight on either DWI$_{hb}$ or T2W lowered the DSC performance, generally consistent with the image availability analysis in \ref{sec:results.interp.availability}. 

More specifically, the best rules for linear mixture model were $\mathbf\upalpha=[0.5,0.5,0]^{\top}$ in the PZ, and $\mathbf\upalpha=[0.4,0.6,0]^{\top}$ in the TZ. What is more, $\mathbf\upalpha=[0.5,0.5,0]^{\top}$ produced the best results in WG as well, due to the size of the PZ being higher in the prostate than the TZ.

\begin{table*}
\caption{Top five rules in terms of pixel-level evaluation metrics via grid search on the validation data of the non-linear stacking model, and tested on the test data. The $No._{rule}$ is decimal conversion of decision $\mathbf{d}^T$.}
\label{tab:gridsearch_voxel}
\resizebox{\textwidth}{!}{
\begin{tabular}{llllllllllllllll}
\hline
{Zone}&{$No.^{rule}$}&\multicolumn{8}{c}{decisions $\textbf{d}^\top$}&\multicolumn{4}{c}{hyperparameters $\mathbf\upbeta$}&{DSC}&{$D_{HD}$}.\\ \hline
\multirow{5}{*}{WG} & 47 & 0 & 0 & 1 & 0 & 1 & 1 & 1 & 1 & 31.51  & 15.34  & -15.49 & -7.74 & 0.33(0.20) & 17.40(8.92) \\
                    & 63 & 0 & 0 & 1 & 1 & 1 & 1 & 1 & 1 & 18.17  & 18.17  & -0.20  & -8.53 & 0.31(0.20) & 17.84(9.39) \\
                    & 76 & 0 & 1 & 0 & 0 & 1 & 1 & 0 & 0 & 15.38  & -31.55 & 15.38  & -7.49 & 0.31(0.19) & 18.24(8.46) \\
                    & 43 & 0 & 0 & 1 & 0 & 1 & 0 & 1 & 1 & 17.53  & 17.53  & -17.74 & -8.58 & 0.31(0.21) & 18.25(9.06) \\
                    & 8  & 0 & 0 & 0 & 0 & 1 & 0 & 0 & 0 & 16.35  & -16.74 & -16.74 & -8.52 & 0.31(0.18) & 17.88(9.32) \\\hline
\multirow{5}{*}{TZ} & 32 & 0 & 0 & 1 & 0 & 0 & 0 & 0 & 0 & -16.74 & 16.35  & -16.74 & -8.52 & 0.16(0.15) & 18.81(8.99) \\
                    & 48 & 0 & 0 & 1 & 1 & 0 & 0 & 0 & 0 & -18.37 & 17.98  & -0.21  & -9.03 & 0.16(0.15) & 18.58(8.67) \\
                    & 34 & 0 & 0 & 1 & 0 & 0 & 0 & 1 & 0 & -0.21  & 17.98  & -18.37 & -9.03 & 0.15(0.14) & 19.11(9.21) \\
                    & 63 & 0 & 0 & 1 & 1 & 1 & 1 & 1 & 1 & 18.17  & 18.17  & -0.20  & -8.53 & 0.15(0.16) & 17.93(8.54) \\
                    & 43 & 0 & 0 & 1 & 0 & 1 & 0 & 1 & 1 & 17.53  & 17.53  & -17.74 & -8.58 & 0.15(0.14) & 18.51(9.01) \\\hline
\multirow{5}{*}{PZ} & 47 & 0 & 0 & 1 & 0 & 1 & 1 & 1 & 1 & 31.51  & 15.34  & -15.49 & -7.74 & 0.37(0.21) & 16.90(9.23) \\
                    & 63 & 0 & 0 & 1 & 1 & 1 & 1 & 1 & 1 & 18.17  & 18.17  & -0.20  & -8.53 & 0.35(0.23) & 17.29(9.65) \\
                    & 76 & 0 & 1 & 0 & 0 & 1 & 1 & 0 & 0 & 15.38  & -31.55 & 15.38  & -7.49 & 0.35(0.20) & 17.70(8.79) \\
                    & 8  & 0 & 0 & 0 & 0 & 1 & 0 & 0 & 0 & 16.35  & -16.74 & -16.74 & -8.52 & 0.35(0.21) & 17.44(9.37) \\
                    & 15 & 0 & 0 & 0 & 0 & 1 & 1 & 1 & 1 & 19.94  & -0.36  & -0.36  & -9.42& 0.35(0.22)  & 17.52(9.32)\\\hline

\end{tabular}}
\end{table*}
\paragraph*{Results of nonlinear decision rules}

 Based on the test set, the top performing rules from the grid search are in 
Table \ref{tab:gridsearch_voxel} and \ref{tab:gridsearch_lesion}, ranked by the voxel-level and lesion-level accuracy, respectively. Rule 47 achieved the best DSC ($0.37\,\pm\,0.21$) and $D_{HD}$ ($16.90\,\pm\,9.23~\mathrm{mm}$) for PZ lesions. Rule 32 and 63 achieved the best DSC ($0.16\,\pm\, 0.15$) and $D_{HD}$ ($17.93\,\pm\, 8.54~\mathrm{mm}$) for TZ lesions, respectively. 
Investigating the hyperparameters sampled for the nonlinear decision rules (\ref{sec:method.hp-estimation.search}), which represent the odds-ratios for individual image modalities, the following conclusions are thus drawn, consistent with those from the linear models.

\begin{itemize}
\item The best lesion localisation results were achieved without ADC in terms of voxel-level results. This conclusion can be evident by 
Table \ref{tab:gridsearch_voxel} that 4, 5 and 4 out of the top five rules are fitted with decisions whose hyperparameters for ADC are negative values in WG, TZ and PZ, respectively. ADC was the least indicative of lesion localisation among the three image modalities.
\item For PZ lesions, according to voxel-level results in 
Table \ref{tab:gridsearch_voxel}, all hyperparameters of the T2W modality are positive in the top five rules. Thus, a lesion was considered positive if it was found positive on T2W, regardless of findings from any other images. The difference in voxel-level accuracy is arguably marginal. 
\item The TZ lesions were considered positive if found positive on DWI$_{hb}$, without considering any other image modalities.
\item In terms of lesion-level metrics, DWI$_{hb}$ and ADC weighed more than T2W according to 
Table \ref{tab:gridsearch_lesion}. The top three rules for $Recall{*}^{GT}$ are most likely achieved with positive hyperparameters of DWI, while the top three rules for $Precision{*}^{Pred}$ had positive hyperparameters for DWI$_{hb}$ and ADC images. 
\end{itemize}

\begin{table*}
\caption{Top three rules in terms of lesion-level evaluation metrics via grid search on the validation data of the non-linear stacking model, and tested on the test data. The $No._{rule}$ is decimal conversion of decisions \textbf{d}}\label{tab:gridsearch_lesion}
\resizebox{\textwidth}{!}{
\begin{tabular}{llllllllllllllllll}
\hline
    {Zone}&{Ranked by}&{$No.^{rule}$}&\multicolumn{8}{c}{decisions $\textbf{d}^\top$}&\multicolumn{4}{c}{hyperparameters $\mathbf\upbeta$}&$Rec.*^{GT}$&$Prec.*^{pd}$&AUC\\ \hline
\multirow{9}{*}{WG} & Recall    
                     & 43  & 0 & 0 & 1 & 0 & 1 & 0 & 1 & 1 & 17.53  & 17.53  & -17.74 & -8.58  & 0.82 & 0.55 & 0.67 \\
                     &           & 34  & 0 & 0 & 1 & 0 & 0 & 0 & 1 & 0 & -0.21  & 17.98  & -18.37 & -9.03  & 0.82 & 0.56 & 0.69 \\
                     &           & 48  & 0 & 0 & 1 & 1 & 0 & 0 & 0 & 0 & -18.37 & 17.98  & -0.21  & -9.03  & 0.81 & 0.56 & 0.68 \\
        
                     \cline{2-18}
                     & Precision & 117 & 0 & 1 & 1 & 1 & 0 & 1 & 0 & 1 & -15.49 & 15.34  & 31.51  & -7.74  & 0.74 & 0.64 & 0.69 \\
                     &           & 19  & 0 & 0 & 0 & 1 & 0 & 0 & 1 & 1 & 14.23  & 29.15  & 14.23  & -36.02 & 0.59 & 0.63 & 0.62 \\
                     &           & 17  & 0 & 0 & 0 & 1 & 0 & 0 & 0 & 1 & -0.35  & 17.05  & 17.05  & -25.57 & 0.74 & 0.62 & 0.67 \\
                     \cline{2-18}
                     & AUC       & 113 & 0 & 1 & 1 & 1 & 0 & 0 & 0 & 1 & -17.74 & 17.53  & 17.53  & -8.58  & 0.81 & 0.61 & 0.71 \\
                     &           & 117 & 0 & 1 & 1 & 1 & 0 & 1 & 0 & 1 & -15.49 & 15.34  & 31.51  & -7.74  & 0.74 & 0.64 & 0.69 \\
                     &           & 34  & 0 & 0 & 1 & 0 & 0 & 0 & 1 & 0 & -0.21  & 17.98  & -18.37 & -9.03  & 0.82 & 0.56 & 0.69 \\
                     \hline
\multirow{9}{*}{TZ} & Recall    & 47  & 0 & 0 & 1 & 0 & 1 & 1 & 1 & 1 & 31.51  & 15.34  & -15.49 & -7.74  & 0.62 & 0.28 & 0.41 \\
                     &           & 48  & 0 & 0 & 1 & 1 & 0 & 0 & 0 & 0 & -18.37 & 17.98  & -0.21  & -9.03  & 0.61 & 0.28 & 0.43 \\
                     &           & 34  & 0 & 0 & 1 & 0 & 0 & 0 & 1 & 0 & -0.21  & 17.98  & -18.37 & -9.03  & 0.59 & 0.29 & 0.44 \\
                     \cline{2-18}
                     & Precision & 19  & 0 & 0 & 0 & 1 & 0 & 0 & 1 & 1 & 14.23  & 29.15  & 14.23  & -36.02 & 0.30 & 0.34 & 0.32 \\
                     &           & 1   & 0 & 0 & 0 & 0 & 0 & 0 & 0 & 1 & 14.74  & 14.74  & 14.74  & -37.27 & 0.34 & 0.32 & 0.34 \\
                     &           & 49  & 0 & 0 & 1 & 1 & 0 & 0 & 0 & 1 & -15.23 & 30.51  & 14.83  & -22.59 & 0.40 & 0.32 & 0.37 \\
                     \cline{2-18}
                     & AUC       & 34  & 0 & 0 & 1 & 0 & 0 & 0 & 1 & 0 & -0.21  & 17.98  & -18.37 & -9.03  & 0.59 & 0.29 & 0.44 \\
                     &           & 48  & 0 & 0 & 1 & 1 & 0 & 0 & 0 & 0 & -18.37 & 17.98  & -0.21  & -9.03  & 0.61 & 0.28 & 0.43 \\
                     &           & 63  & 0 & 0 & 1 & 1 & 1 & 1 & 1 & 1 & 18.17  & 18.17  & -0.20  & -8.53  & 0.56 & 0.28 & 0.43 \\
                     \hline
\multirow{9}{*}{PZ} & Recall    & 34  & 0 & 0 & 1 & 0 & 0 & 0 & 1 & 0 & -0.21  & 17.98  & -18.37 & -9.03  & 0.76 & 0.64 & 0.69 \\
                     &           & 113 & 0 & 1 & 1 & 1 & 0 & 0 & 0 & 1 & -17.74 & 17.53  & 17.53  & -8.58  & 0.75 & 0.67 & 0.70 \\
                     &           & 63  & 0 & 0 & 1 & 1 & 1 & 1 & 1 & 1 & 18.17  & 18.17  & -0.20  & -8.53  & 0.73 & 0.59 & 0.67 \\
                     \cline{2-18}
                     & Precision & 17  & 0 & 0 & 0 & 1 & 0 & 0 & 0 & 1 & -0.35  & 17.05  & 17.05  & -25.57 & 0.66 & 0.68 & 0.66 \\
                     &           & 113 & 0 & 1 & 1 & 1 & 0 & 0 & 0 & 1 & -17.74 & 17.53  & 17.53  & -8.58  & 0.75 & 0.67 & 0.70 \\
                     &           & 117 & 0 & 1 & 1 & 1 & 0 & 1 & 0 & 1 & -15.49 & 15.34  & 31.51  & -7.74  & 0.67 & 0.67 & 0.67 \\
                    \cline{2-18}
                     & AUC       & 113 & 0 & 1 & 1 & 1 & 0 & 0 & 0 & 1 & -17.74 & 17.53  & 17.53  & -8.58  & 0.75 & 0.67 & 0.70 \\
                     &           & 34  & 0 & 0 & 1 & 0 & 0 & 0 & 1 & 0 & -0.21  & 17.98  & -18.37 & -9.03  & 0.76 & 0.64 & 0.69 \\
                     &           & 43  & 0 & 0 & 1 & 0 & 1 & 0 & 1 & 1 & 17.53  & 17.53  & -17.74 & -8.58  & 0.70 & 0.62 & 0.68 \\
\hline
\end{tabular}}
\end{table*}

\begin{figure*}
\caption{The results of nonlinear stacking, linear mixing, baseline models, and base networks are visualised from left to right, respectively. The first block and the second block show DSC results of peripheral zone lesion and transition zone lesion, respectively. The third block shows the lesion-level results ($Precision{*}^{Pred}$ and $Recall{*}^{GT}$, respectively) in the brackets. {The red and green contours represent ground truth and prediction, respectively\label{fig:visual}}}
\includegraphics[width=\linewidth]{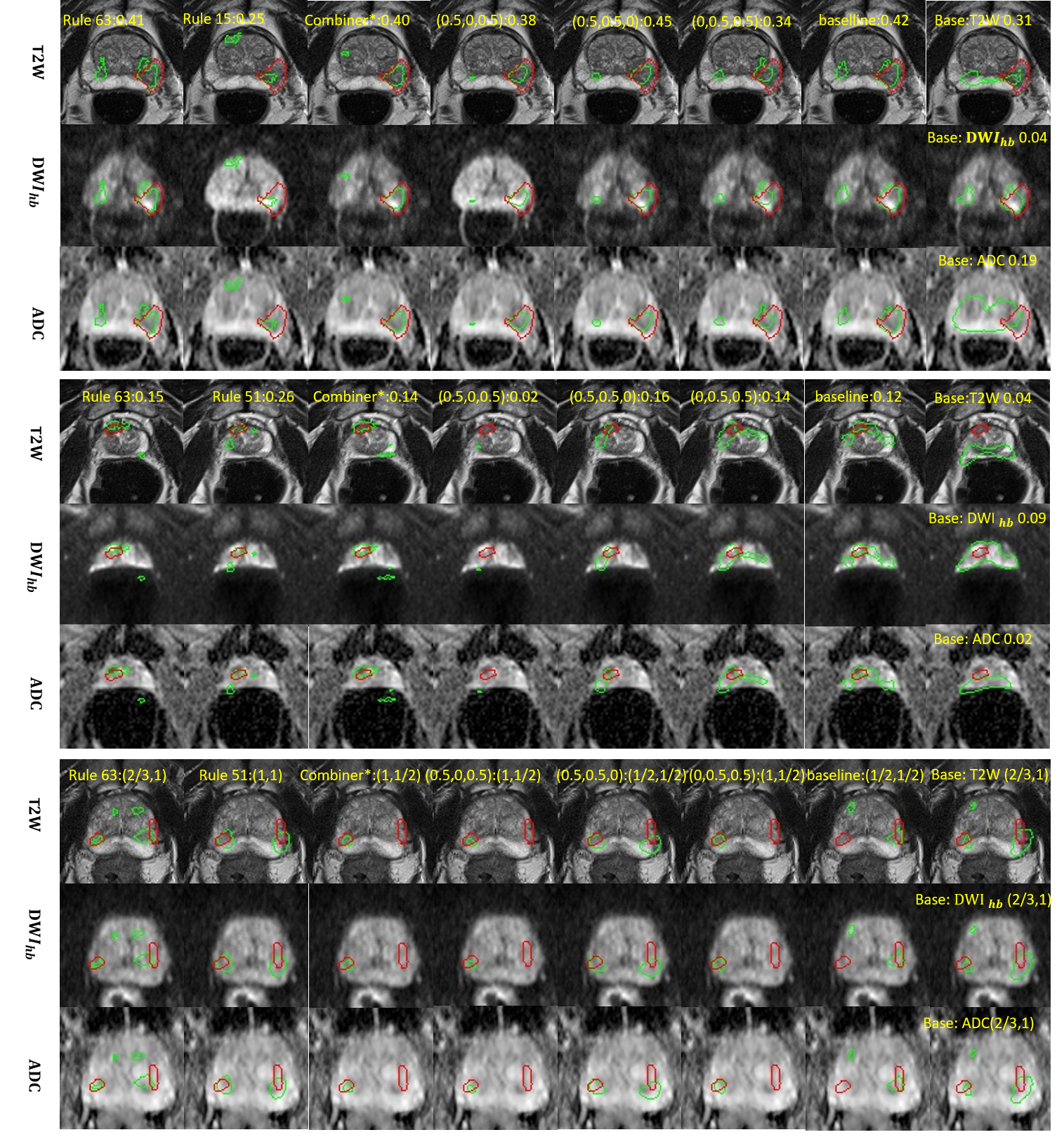}
\end{figure*}

\subsubsection{Clinical case studies}

 \ref{fig:visual} shows two examples of test cases with segmentation results from different methods. Rule 63 $\mathbf{d}=[0,0,1,1,1,1,1,1]^{\top}$ is comparable to the baseline method, which emphasises T2W and ADC prediction during the combining model, yielding a DSC of 0.41. In contrast, Rule 15, which emphasises only T2W in the combining model, has a much lower DSC than Rule 63 (0.25 vs. 0.41). As we stated in the previous section, the T2W$+$DWI$_{hb}$ combination improves the accuracy of lesion localisation. The third column is predicted by the non-linear stacking model of the Combiner model (denoted Combiner*), which has a DSC comparable to Rule 63.

In terms of the linear mixture model, the combination T2W$+$DWI$_{hb}$ produces the best DSC of 0.45. The variation of results with different linear combinations is not as high as the non-linear stacking model. The final column shows the results for the Base networks with single images, in which the DWI produces the best DSC, but lower than the baseline approach.

For TZ lesions from the 4th to 6th rows, the non-linear stacking model with Rule 51 $\mathbf{d}=[0,0,1,1,0,0,1,1]^{\top}$, emphasising DWI in the combined model, achieved the best DSC of 0.26 among all methods, much higher than the baseline method. In addition, the linear mixed model of T2W$+$ADC produced the worst segmentation mask compared to the other two combinations. By comparing the linear mixed model with the baseline network, we conclude that multiple types of images indeed benefit the localisation of lesions.

\section{Discussion}

\label{sec:discussion}
In clinical decision-making, it is common practice to combine multiple sources of information to reach an informed consensus. An example of this is seen in uro-radiology, localising prostate cancer from multiple types of MR images. Many recently developed deep learning methods focus on exploiting the representation ability of neural networks without considering how different image modalities from mpMR images are combined. This work investigates approaches that model the combining process effectively and efficiently without sacrificing predictive ability. The proposed Combiner and HyperCombiner networks bring explainable, quantitative decision-making for multimodal inputs and enhance the transparency and interpretability of how they are combined. 

The proposed approach allows the comparison between different image modalities, whose availability is associated with real-world cost and feasibility. As a result of this work, we have proposed decision rules that may be advantageous
for emerging machine learning models, but these may also be considered new potential protocols for further radiological practice. What is not included in this work is the ability to incorporate, using the proposed method, known prior
knowledge that is related to the importance of individual types, such as quantifiable quality of image quality~\citep{saeed2022image} and local
expertise. For example, given a local estimate of below-average image quality and resonance time constraints, certain image modalities may need to be weighed down to reach optimum tumour localisation performance. 

Future work may consider multiple classes of tumours, such as those based on the five-point PI-RADS scores. In addition to the intended radiology application, the provided analysis on the image modality availability may offer valuable insights to urologists. It provides indications regarding the modalities that hold the most value when facing choices such as repeated biopsy, therapy options or surgery. This may also benefit from a system that is learned using the histopathology labels when they are available. However, developing better-than-radiologist machine-learning models learned from histopathology labels remains an open research question.
 Furthermore, future work can also focus on exploring the utilisation of lesion size in conjunction with other performance-impacting factors, including image quality, lesion grading, irregularity, heterogeneity, and lesion location, that can potentially improve accuracy.

More generally, the proposed approach opens an investigation between the so-called ``black-box'' approaches represented by deep learning and the classical decision rule modelling. In particular, biomedical imaging applications have seen an abundance of multimodal input tasks. In this context, the hybrid machine learning framework proposed can provide benefits and improvements.

The results presented in this work need to be interpreted with their limitations. For example, although this study used a relatively large data set from real clinical patients, the imaging and label data come from a single referral centre with significant experience in imaging and mpMR reading. Multi-centre research may be required to generalise specific clinical conclusions further. For example, the optimised rules may be subject to local protocols; therefore, further constraints or characteristics should be taken into account. When these conditions are identified, the proposed decision rule modelling approach should also be applicable.
 
The proposed method also has the potential to be applicable to other clinical applications, such as patient-level classification and image quality assessment. However, it is important to clarify that evaluating its performance in this context may require different types of networks and patient cohorts, such as the screening population, rather than the focal therapy patients used in this study. 

\section{Conclusion}

PI-RADS rule has widely been used in lesion localisation by radiologists. Based on this rule, we proposed Combiner networks that use pre-defined decision rules to combine the modality-specific class probabilities using either linear mixture model or nonlinear stacking model. To further facilitate the capability of Combiner networks, we have developed HyperCombiner networks, allowing for more efficient hyperparameter analysis. The applications of the resulting Combiner and HyperCombiner networks are demonstrated using a sizeable clinical mpMR data set. We presented extensive experimental results to show a number of interesting comparisons between different image modalities and a potential to devise new and more effective rules to combine the modalities under the specific clinical context of localising prostate cancer.

\section*{Acknowledgment}
This work was supported by the International Alliance for Cancer Early Detection, an alliance between Cancer Research UK [C28070/A30912; C73666/A31378], Canary Center at Stanford University, the University of Cambridge, OHSU Knight Cancer Institute, University College London and the University of Manchester. This work was also supported by the Wellcome/EPSRC Centre for Interventional and Surgical Sciences [203145Z/16/Z]. For the purpose of Open Access, the author has applied a CC BY public copyright licence to any Author Accepted Manuscript version arising from this submission.

\appendix{}  
\section[pfx={Appendix},nonum]{Supplementary Material}

\subsection{Derivation of decision rules based on { {PI}}-{ {RADS}}}

\label{sec:app.rules}

PI-RADS is a guideline designed for standardising radiologist reports of detecting and grading clinically significant prostate cancer on mpMR images. It aims to streamline the use of three MR modalities of T2W, DWI and DCE, with which DWI may be examined with one or more image modalities such as DWI$_{hb}$ and ADC maps. DCE is considered if other clear practical benefits do not outweigh its diagnostic value. A key feature of PI-RADS is that it has constantly been updated in response to emerging evidence. While it is out of the scope to discuss the practical use of the current version of PI-RADS, a binary classification version is derived from it for the purpose of demonstrating the proposed machine learning methodology.

To summarise, the current version of PI-RADS first defines dominant modality, T2W and DWI, for lesions found in TZ and PZ, respectively, before modifying the final grading, on a scale of 1--5, based on other image modalities. This is illustrated in the left diagram in \ref{fig:pirads} (left). As an example, considering a positive lesion with a score greater than or equal to 3 -- negative otherwise -- the binary classification used in this study is illustrated in the right diagram in \ref{fig:pirads} (right). It is noteworthy that PI-RADS does not explicitly distinguish DWI$_{hb}$ and ADC, since these are both diffusion-based modalities. Therefore, there has not been wide agreement when the presentation of disease differs between the two image modalities. For the purpose of quantifying between these two image types, this study considers either DWI$_{hb}$ or ADC being greater than or equal to 3 is equivalent to DWI being greater than or equal to 3, while for transition zone lesions, it needs both DWI$_{hb}$ and ADC being greater than or equal to 3 in order to upgrade a negative lesion to a positive one. 
Table \ref{tab:condition_vectors_pz} summarises the condition and decision of lesion in peripheral zone.

\begin{table}[!hb]
    \centering
    \caption{Example condition vectors and decisions for peripheral zone (PZ) lesions, $\tau$=1,2 and 3 represent T2W, DWI$_{hb}$ and ADC, respectively. } \label{tab:condition_vectors_pz}
    \resizebox{0.5\textwidth}{!}
    {
    \begin{tabular}{ |c|c|c|c|c| } 
    \hline
     $\mathcal{C}$ & $p(\mathcal{C}\mid\tau=1)$ & $p(\mathcal{C}\mid\tau=2)$ & $p(\mathcal{C}\mid\tau=3)$ & $p(\mathcal{C})$ \\ 
    \hline
    PZ & negative & negative & negative & NEGATIVE \\ 
    PZ & negative & negative & positive & POSITIVE \\ 
    PZ & negative & positive & negative & POSITIVE \\ 
    PZ & negative & positive & positive & POSITIVE \\
    PZ & positive & negative & negative & NEGATIVE \\ 
    PZ & positive & negative & positive & POSITIVE \\ 
    PZ & positive & positive & negative & POSITIVE \\ 
    PZ & positive & positive & positive & POSITIVE \\ 
    \hline
    \end{tabular}
    }
\end{table}

\begin{equation}
\begin{aligned}
  \textbf{R} =
  \left[
    \begin{array}{cccccccc}
     0 & 0 & 0 & 0 & 1 & 1 & 1 & 1\\
     0 & 0 & 1 & 1 & 0 & 0 & 1 & 1\\
     0 & 1 & 0 & 1 & 0 & 1 & 0 & 1\\
    \end{array}
  \right], 
  \textbf{d} = 
  \left[
    \begin{array}{c}
     0\\ 1\\ 1\\ 1\\ 0\\ 1\\ 1\\ 1\\ 
    \end{array}
  \right]
  \end{aligned}
  \label{eq:condition-vectpz}
  \end{equation}
where $\textbf{R}=[\textbf{r}_1,...,\textbf{r}_K]$ and $\textbf{d}=[d_1,...,d_K]^\top$.

Given the constant condition vectors \textbf{R} (Eq.~\eqref{eq:condition-vectpz}), the resulted decisions \textbf{d}$^{(TZ)}$, \textbf{d}$^{(PZ)}$ and \textbf{d}$^{(WG)}$, are summarised in 
Table \ref{tab:results_pirads_fit}, for TZ, PZ and WG lesions, respectively. The fitted hyperparameters, $\hat{\mathbf\upalpha}$ and $\hat{\mathbf\upbeta}$, can also be found in 
Table \ref{tab:results_pirads_fit}, for a linear stacking model and a nonlinear stacking model, respectively, using methods described in \ref{sec:method.hp-estimation.PI-RADS}.

\subsection{The acceptance--rejection algorithm for non-linear stacking model hyperparameter sampling}

\label{sec:app.alg}
Since unnecessarily extreme values of the defined hyperparameters can lead to diverged optimisation during training, we determine the range of hyperparameters $\upbeta$ by solving the Eq.~\eqref{eq:encode.nonls} in a least-square sense with sampled all possible conditions and decisions. 
An acceptance--rejection algorithm was followed to indirectly sample hyperparameter values for the nonlinear stacking models, as outlined in  Algorithm \ref{alg:nonlinear}. 
In this application, these sampled values were used as hyperparameters $\upbeta$ for the training of HyperCombiner networks with the nonlinear stacking model. 
A set of $N$ samples were pre-computed for being randomly sampled in each training iteration. During inference, the grid search was conducted among all these hyperparameter values for rule discovery. 

\begin{algorithm}
    \caption{Rejection sampling for nonlinear stacking model}    \label{alg:nonlinear}
    \SetAlgoLined
        \KwData {Required sample size $N=2^8$}
        \KwResult {Sampled hyperparameter set $\{\mathbf\upbeta_1, ..., \mathbf\upbeta_N\}$ }
        Initialise counter $n=1$, a constant condition vector set $\textbf{R}$\\
    \While{$n \leq N$}
        {
        \textbf{STEP1:}
            Convert n to an 8-bit binary number as $\textbf{d}_n$ \\
        \textbf{STEP2:}
            Compute for $\mathbf\upbeta_n$ \\
        \For{i=0, $i++$, $i\leq 10^4$}
            {           $\hat{\textbf{d}}_n=\mathbf\upsigma([\textbf{R}^\top,\textbf{1}] \cdot \mathbf\upbeta_n)$\\
            $\textit{loss}=-\sum{p(d_{k,n}) \cdot \log_{}p(\hat{d}_{k,n})}$\\
            $\mathbf\upbeta_n \longleftarrow \mathbf\upbeta_n + \nabla \textit{l}(\mathbf\upbeta_n)$
            }
        \textbf{STEP3:}\\
        \If {$ \|\textbf{d}_n-\hat{\textbf{d}}_n\|_2^2 \leq \frac{\eta^2}{8}$\footnotemark{}}
         { n++}
    }
\end{algorithm}
\footnotetext{$\eta$ is a small number that makes sure the l2 difference between \text{d} and $\hat{\textbf{d}}$ is same, here, $\eta=0.5$.}

\bibliographystyle{unsrtnat}
\bibliography{refs}
\clearpage
\end{document}